\documentclass[journal=jacsat,manuscript=article]{achemso}

\usepackage[version=3]{mhchem} 
\usepackage{upgreek,textgreek}

\author{Anjana E Sudheer}
\author{Golla Tejaswini}
\author{D Murali}
\affiliation
{Indian Institute of Information Technology Design and Manufacturing Kurnool, 518007, Andhra Pradesh, India}
\email{dmurali@iiitk.ac.in}
\author{Matthias Posselt}
\affiliation
{Helmholtz-Zentrum Dresden-Rossendorf, Bautzner Landstrasse 400, 01328 Dresden, Germany}

\title
  {First Principles Investigation on Structural and Optoelectronic Properties of newly designed Janus Lead Halides PbXY (X, Y = F, Cl, Br, I )}

\begin{document}


\begin{abstract}
Inspired by the beauty in the asymmetry, we design a novel class of Janus structures PbXY (X,Y = F, Cl, Br, I) and propose it for the solar mediated photocatalytic water splitting hydrogen production as well as for the photovoltaic solarcell applications.
These novel Janus structures show large modulation in layer thickness, bond lengths and bond angles due to asymmetry of two sides. Charge analysis shows that covalent bonding for less electronegative atoms (I and Br) and ionic bonding for more electronegative (Cl and F) atoms. Strong dual bonding like ionic one side and covalent other side is observed when heavy and lighter atoms are part of the same Janus structures. The as designed Janus structures show good dynamical stability through phonon calculations. Basic electronic structure using Generalised Gradient Approximation (GGA) reveal both direct and indirect nature of band gap with large tunability varying from 2.5 to 3.5 eV. Such a large tunability of band gap can be exploited for multifunctional applications. Heyd-Scuseria-Ernzerhof (HSE) electronic structure calculations are performed for more accuracy and wider band gaps are predicted for these Janus structures. The calculated electron and hole effective masses show robust charge carrier dynamics. The orbital resolved electronic  density of states (DOS) shows that the conduction band edge is composed of pz orbital of Pb atom. The partial charge density calculated at conduction band minimum (CBM) also support the result obtained from PDOS analysis.  Breaking of centrosymmetry, covalent bonding along z-direction, polarization in the out of plane direction, the z-oriented orbitals of CBM all points that these materials are suitable for shift current generation. The calculated optical absorption spectra show that the Janus structures are suitable for visible light absorption. The calculated potential difference between the top and bottom layer show significant variation and maximum (1.02 eV) is observed for PbClF. Further, we show that combining both potential difference and HSE bandgap, valence band maximum (VBM) and CBM straddle the water redox potentials, thus making the Janus structures suitable for hydrogen evolution reaction (HER) and oxygen evolution reactions (OER) on the opposite sides of the Janus structures.
\end{abstract}

\section{Introduction}
'Janus', the word remind us Roman God, symbolised by the two headed image facing in opposite direction. As the image suggests, this indicate union of two entirely contradictory states such as past and future, beginning and ending, war and peace. But, the research field started to notice Janus from 1990s, after the usage of word Janus by Nobel laureate de Gennes\cite{1_1}. The early papers discuss Janus particles which was known by the amalgamation of two opposite behaviours, namely hydrophilicity on one side and hydrophobicity on the other side\cite{2}. But, the Janus material got much attention and research interest in the recent years with the arise of two dimensional (2D) Janus materials\cite{29,30}.\\
The first proposed 2D Janus structures are based on graphene structure\cite{4_1}.  'Graphone', the hydrogenated graphene in 2009\cite{4_2} and later halogenated graphene in 2014  were predicted using theoretical simulation techniques\cite{4_3}. After the invention of graphene based Janus, the research expanded to the modification of different transition metal dichalcogenides(TMDs) such as MoS$_2$, WS$_2$, PtS$_2$, and SnS$_2$ monolayers for the development of Janus 2D materials. By the modification of the above mentioned TMDs, the novel Janus structures MoXY\cite{4}, WS$_2$, PtS$_2$, and SnS$_2$ monolayers for the development of Janus 2D materials. By the modification of the above mentioned TMDs, the novel Janus structures MoXY\cite{4}, WXY\cite{5,32,55}, PtXY\cite{6,33,54} and SnXY (X,Y = S, Se,Te) \cite{6_1,34} were successfully predicted using Density Functional Theory (DFT) simulations. The as predicted Janus TMDs are reported to be efficient for photocatalytic as well as photovoltaic applications. In 2021, the designing of Janus structure HfSO using DFT simulations was successful by the research group lead by the Vo D. Dat et al.\cite{56}.
This material was proposed for different areas of research such as photocatalysis, photovoltaics and in the field of optoelectronics as well. Further, the Janus structure which got much attention in the field of energy material research was Pb$_2$SSe designed by Fuzheng Zang et al.\cite{57}. In general, the application of these Janus  structures were not limited to catalysis \cite{7_1}and photovoltaics\cite{7_2}, they were explored in the thermoelectrics\cite{58,9}, optoelectronics\cite{8,37} and also for the multifunctional device\cite{9_1} applications.
Motivated from the unique properties of these Janus structures, research was directed to explore Janus transition metal dihalides. In 2022, Yusheng Hou et al.\cite{11} and Zhenning Sun et al.\cite{11a} discovered Cr based Janus transition metal halides. The primary intention was to explore the magnetic properties of the material. Recent study on new Janus material YMN (Y= Yttrium, M = I, Br and N = Cl, Br) aimed at energy applications, mainly in the field of optoelectronics \cite{12}. The TMDs are already established in the field of energy application as well as  multifunctional devices. By understanding the different possible applications of TMDs, it will be interesting to design  and explore lead based halide structures, owing to the unique properties buried in the prototypical structure. The next-generation energy applications can be improved with the Janus lead halides. Till now, Pb based Janus structures are not yet reported. The only reported two dimensional lead dihalide monolayer is PbI$_2$ monolayer, which is explored for the thermoelectric applications\cite{70}.\\
The Janus structures are actively being pursued in the field of photocatalysis, because they are reported to be materials having visible light active bandgap, low binding energy excitons, longtime spatial charge separation, fast carrier mobility, which is favourable for the water redox reactions. Moreover, they can overcome the major drawback of the existing prototype photocatalysts, which is reported to be fast charge recombination. The construction of Janus is less complex compared to the photocatalyst based on heterostructure \cite{68,69}. By breaking the symmetry of the two dimensional structure, aim is to create polarity in the material\cite{64}. The created polarity inside the material can bring out built in electric field, which will be beneficial for the successful charge separation inside the single material itself. Moreover, the created electric field reduces the charge recombination effectively. By reducing the charge recombination, the life time of the separated charge carriers can be improved. Subsequently, there will be enhancement in carrier dynamics which lead to improvement in catalytic efficiency. \\
The built-in noncentrosymmetric nature of the Janus structure, helps to diversify the application of the material from the field of catalysis to other energy material research. Referring to this non centrosymmetric nature, mainly they are applied in the bulk photovoltaic devices based on the shift current mechanism\cite{65a}. The importance of bulk photovoltaic devices over the normal photovoltaic devices is that it can generate electric field inside a single bulk system itself\cite{61}. So they can produce photocurrent without using the conventional methods based on the heterostructure construction. Generally, the perovskite crystal structures are found to be showing these bulk photovoltaic effect due to the polar nature and the lack of inversion symmetry\cite{77}. Other than perovskite structures, the state of the art research is going on mainly on the Janus monolayered structures which show inversion symmetry breaking, due to the difference in top and bottom layers\cite{77_a}. They are widely explored for the solar cell devices, as it can produce open circuit photovoltages larger than their bandgap energies\cite{79}. Apart from the non-centrosymmetry, shift current is enhanced by the covalent bonding and favourable orbital character in the band edges\cite{74}.\\
In short, we can conclude that the non-centrosymmetric Janus structures are good for  photovoltaic devices as well as  photocatalytic materials due to the out of plane polarisation and the built-in electric field inside the material. Owing to the above mentioned special features, they can also be proposed for the  study of piezoelectricity, and the ferroelectricity\cite{62,75,77}, considering that these are the general properties exhibited by the non-centrosymmetric structures. Nowadays, the synergy of different properties in single material is being explored for varied applications. Recently, novel structure MZX (M = Ga, In; Z=Si, Ge ;X = S, Se, Te) monolayers were explored for photocatalytic, thermoelectric and for the topological properties\cite{78}. In addition to that, 2D materials are really interesting systems for the storage of quantum information due to the reduced dimensionality. This lead to charge confinement and hence many unique quantum states that are extremely stable\cite{10}. Eventhough the Janus structures have this wide variety of applications, the challenge in this field is to design Janus monolayer experimentally. Highly controlled growth of Janus monolayer MoSSe with chemical vapour deposition technique was reported recently by Chan wook Jang et al. \cite{71}. The success of the experimental synthesis of Janus monolayer gives hope that these method can be applied to other Janus structures too.\\
Ultimately, the bird's eye view of this project is to design the structurally and thermodynamically stable Pb based Janus structures and explore for the solar water splitting hydrogen production and photovoltaic applications. As a first step, we designed classic PbX$_2$ monolayers (X = F, Cl, Br, I), which are considered as parent materials for the Janus structures. The detailed results of parent structures are given in supplementary material and only salient features of parent materials are discussed in main text. The present paper is structured as follows: In the first part we discuss the  design of novel Janus structures and discuss the structural parameters and bonding nature through charge analysis. 
The formation energies are also calculated and  discussed in detail. In the third part, the basic electronic properties are studied using both GGA and HSE functionals.  Later, the first order optical property analysis is performed for the different Janus structures using  density functional perturbation theory (DFPT). Next, the band alignment of the Janus structures are calculated using electrostatic potentials and discussed in comparison with the water redox potentials. Finally, the summary of our research work and the outlook are presented.\\
\section{Computational Details}
The \textit{ab initio} DFT calculations are performed using Vienna Ab initio Simulation Package (VASP)\cite{13,38}. The exchange-correlation functionals used for the calculation are Generalised Gradient Approximation-Perdew Burke Ernzerhof (GGA-PBE) functional\cite{14}. Here, better computational calculation efficiency is achieved by treating core and valence electron interactions using Projected Augmented Wave (PAW) method\cite{15,39}. As a first step, the geometrical structures are relaxed using the plane wave basis set with an energy cut off 650 eV. The self iteration scheme used for this geometrical optimization is based on blocked-davidson algorithm.  The forces on each atom are allowed to decrease below 0.001 eV/{\AA} and energy is converged to $1 \times 10^{-6}$  eV. The Brillouin zone sampling  is done with Gamma centered kpoints 16x16x1 for the unit cells\cite{16}. After the complete relaxation, band structure is  calculated along the high symmetry point in the Brillouin zone. For the more accurate electronic structure, the GGA level band diagrams are corrected using the HSE hybrid functional\cite{81}. The frequency dependent dielectric constants are calculated using density functional perturbation theory (DFPT) as implemented in VASP code\cite{76}. VASPKIT interface software is used for extracting data from the vasp output files\cite{17}. The electrostatic potentials are calculated using ionic and Hartree potential. For the geometry visualisation of the structures, VESTA software is used\cite{40}. \\
Further, the dynamical stabilities of the designed Janus structures are verified using the phonon dispersion calculated with the help of phonopy code \cite{18,19}interfaced with VASP. In order to understand the convergence of phonon vibrational spectra with respect to supercell size, the calculations are performed with supercell size 3x3x1 to 7x7x1 with suitable kpoints. The details of the supercell size, kpoints and the error percentage calculated are given in the supplementary details. 
\section{Results and discussions}
\subsection{Structural and Geometrical properties}
For the systematic analysis, we have investigated four different configurations of Pb based dihalide monolayers with chemical formula PbX$_2$ (X = F, Cl, Br, I), which are considered as the parent materials of newly constructed lead based Janus dihalide monolayers. The top  and side view of the geometrically relaxed PbX$_2$ monolayers are given in Fig. S1. From the side view of these PbX$_2$ structures, it is clear that they are trilayered structures with  Pb atoms sandwiched between similar halide layers. The atomic arrangement  of these monolayers are similar to the 1T MoS$_2$ like structure with Pb atom instead of Mo atom and halides instead of sulfur atom. The space group and point group of these PbX$_2$ monolayers are found to be P-3m1(164) and D3d respectively. In the further analysis, the variation of lattice parameters and bond length of these PbX$_2$ monolayers are compared with  atomic size of the halide atoms in PbX$_2$ structures. For that, detailed structural information of these parent monolayers are provided in table S1. The lattice parameters obtained for PbI$_2$ is matching with the experimental results \cite{80}.\\
In PbX$_2$ monolayers, The lattice parameters and bond lengths are highest for PbI$_2$ and lowest for PbF$_2$. When the 'X' atoms in PbX$_2$ monolayer move from I to F, the lattice parameters and bond lengths are found to be gradually decreasing. This is due to the decrease in atomic radius of X atom.  In the further analysis, we found the bond angle that create at the center layer, also decreasing when X atoms in PbX$_2$ structures are changing from  I to F. This suggests that the layer thickness have strong dependence on the atomic radius of atoms present. From the Table S1, PbI$_2$ monolayer is observed to be comparatively thick layer with thickness 3.82 {\AA} and PbF$_2$ is the atomically thin layer with thickness 1.98 {\AA}.  The strong tunability of structural parameters such as lattice constant, bond angle, and layer thickness may enable these materials for diverse applications. \\
\begin{figure*}
\centering
\includegraphics[ height= 20cm]{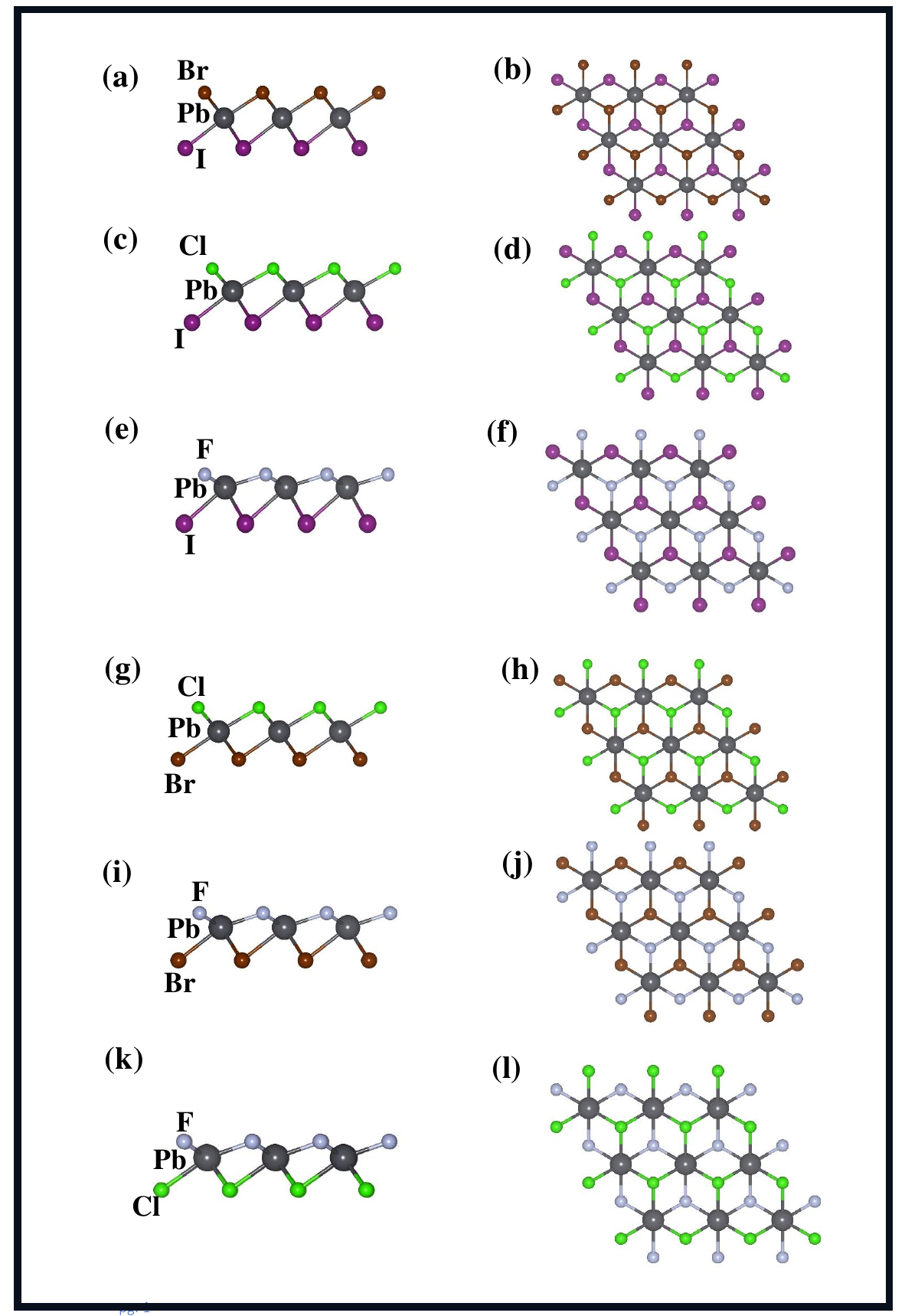}
\caption{The Side and top view of geometrically relaxed structures of PbIBr((a)and (b)),PbICl((c)and(d)), PbIF ((e) and (f)), PbClBr ((g) and (h)), PbFBr ((i) and (j)), PbClF ((k) and (l)). The black, violet, brown, green and grey colours represent Pb, I, Br, Cl and F atoms respectively.}
\end{figure*}
  
Later, from the constructed PbX$_2$ monolayers, we designed novel Pb based Janus structures with chemical formula PbXY (X, Y = F, Cl, Br, I). By replacing one of the halide layer with different halide atoms, we could successfully break the symmetry of PbX$_2$ structures in the out of plane direction. In these structures, Pb atoms are sandwiched in between halide layers with different electronegativity.  We have also noticed that the novel Janus structures have similar PbX$_2$ atomic arrangement. The geometrically relaxed structures of newly designed Janus monolayers are given in Fig. 1. It was noticed that the newly constructed stable structures have trigonal crystal structure with C3v point group symmetry and P3m1 space group symmetry. By analysing the side view of the constructed PbXY monolayers, we have noticed that these Janus structures can be uniquely identified by their signature shift in the top and bottom layer. The shift in different halide layers of the Janus material increases with electronegativity of the atoms present in the Janus structures. Further, the structural parameters such as lattice parameter, bond length, bond angle and the layer thickness are calculated and presented in Table 1. We noticed that every PbXY Janus structures show lattice parameters in between the parent materials of PbX$_2$ and PbY$_2$ monolayers. Such a similar trend was previously reported in case of Sn based Janus structures\cite{67}. In the lead iodide structures (PbIX (X = Br ,Cl ,F)), the trend of decreasing lattice parameter is noticed with increasing electronegativity of the other layer.  
\begin{table*}
    \caption{Structural information of PbXY (X,Y = F, Cl, Br, I) Janus structures.}
    \scriptsize 
\renewcommand{\arraystretch}{1}
    \begin{tabular}{ |p{2cm}|p{1.9cm}|p{4.5cm}| p{2cm}|p{2cm}|}   
    \hline
\normalsize{Structure} &\normalsize{Lattice parameter (\AA) (a=b) }&\normalsize{Bond length (\AA)}& \normalsize{Bond angle (Degree) (@centerlayer)} & \normalsize{Layer thickness (\AA) }\\[0.5ex]
    \hline\hline
    \hspace{0.5cm}
    \vspace{0.25 cm}
\small PbIBr &	\small4.49 &\small	 3.24(Pb-I), 3.06(Pb-Br)&	\small89.12	& \small3.6\\
\hspace{0.5cm}
\vspace{0.25cm}
\small PbICl &\small	4.45 &	 \small 3.24(Pb-I), 2.93(Pb-Cl)	& \small 87.01&\small 3.4\\
\hspace{0.5cm}
\vspace{0.25cm}
\small PbIF &\small 4.24 & \small3.21(Pb-I), 2.55(Pb-F) & \small79.82 & \small2.83 \\[1ex]
\hspace{0.5cm}
\vspace{0.25cm}
\small PbBrCl & \small4.40 & \small3.06(Pb-Br), 2.93(Pb-Cl) & \small85.49 &\small 3.19 \\[1ex]
\hspace{0.5cm}
\vspace{0.25cm}
\small PbBrF &\small 4.15 & \small 3.03(Pb-Br), 2.52(Pb-F)&\small 79.56 &\small 2.66  \\[1ex]
\hspace{0.5cm}
\vspace{0.25cm}
\small PbClF & \small4.12&\small 2.90(Pb-Cl), 2.52 (Pb-F) &\small 45.69 &\small 2.5 \\[1ex]
\hline
\end{tabular}
\end{table*}
The same trend is noticed in case of bond length, bond angle at the center layer and the layer thickness of the different PbIX Janus structures. Also, the lead bromide structures PbBrX (X = Cl ,F) shows the trend of decreasing structural parameters when the 'X' layer is occupied with more electronegative atom. Due to the difference in atomic radius of different halides, the bond length between Pb-X and Pb-Y are different in the PbXY Janus structures. We found that larger the atomic radius of halide atom in a Janus structure, larger will be the bond length with Pb atom. In case of PbIBr and PbICl, the Pb-I bond length is same, indicating that the presence of other halide layer does not affect Pb-I chemical environment. 
But, Pb-I bond in the PbIF structure is varied from that of PbIBr, indicating that the highly electronegative fluorine atom could change the chemical environment of Pb-I bond. Similarly, for other janus structures also, one halide layer has influence on the position of other side halide layer. \\
To understand the charge transfer inside the material, differential charge density (DCD) of PbX$_2$ parent materials (Fig. S2) and PbXY Janus structures (Fig. 2) are analysed. The differential charge density is calculated by substracting the sum of the individual atomic charge densities from the total charge density of the relaxed structure.
It is found that, there is a depletion of negative charge from Pb and accumulation of the negative charge in the top and bottom halide layers in all the structures. The charge accumulation in the halide layers of PbX$_2$ monolayers are found to be nearly dumb bell shaped except in the PbF$_2$ monolayer (Fig. S2 (d)). Near to fluorine the charges accumulated is  completely in spherical shape. From the analysis of Fig. 2, the charge accumulation is observed to be high around the more electronegative halide layers in each Janus structures. Similar to the PbX$_2$ monolayers, the charge accumulated region near fluorine in the Janus structures is showing completely  spherical in shape and for the other halides with nearly in dumb bell shape. This shows that the F is more ionic compared to the other halide atoms.\\
\begin{figure*}[h]
\centering \includegraphics[height= 8 cm]{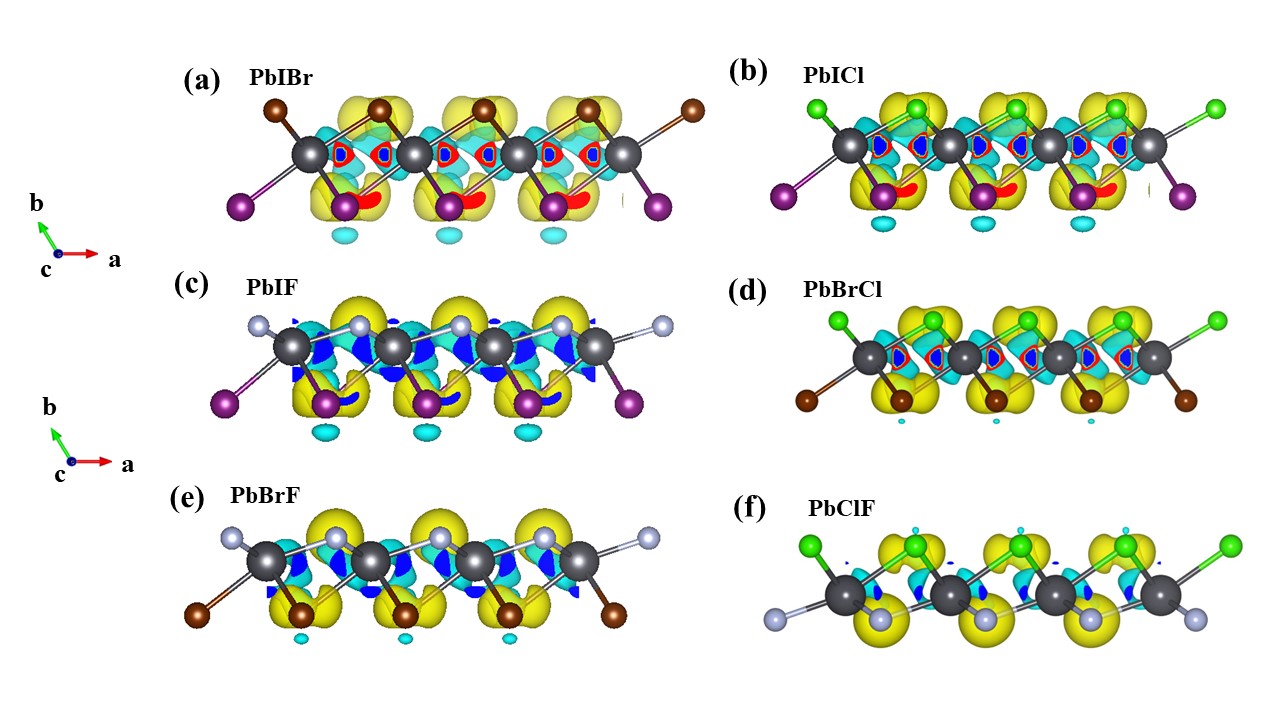}
 \caption{Differential charge density representation of  PbIBr (a), PbICl (b), PbIF (c), PbBrCl (d), PbBrF (e), and PbClF (f) monolayers with isosurface level 0.002 e/{\AA$^3$}. The yellow and blue represent charge accumulated and depleted region respectively. The charge accumulation observed to be in spherical shape near to more electronegative atom and dumbbell shaped near to the less electronegative atom.}
\end{figure*} 

For the quantitative analysis of charge transfer, we have performed the Bader charge analysis on the different PbX$_2$ monolayers (Table S2) and the newly constructed Janus structures (Table 2). From the calculated Bader charges on each element in the monolayers, we define the following formula:
\begin{equation}
      \Large \Delta Q = {Z}_{valence}-BA
\end{equation}
where,  \(\Delta Q\) is the charge depleted or accumulated from the chosen atom. 
The \({Z}_{valence}\) and BA are the number of valence electrons and number of electrons determined by the Bader analysis respectively.
The nominal charge state of Pb and other halides are 2+ and  1- respectively. The value \(\Delta Q\) is compared with the nominal charge states of Pb and halide atoms, which are presented in Table 2. If the difference between \(\Delta Q\) and the nominal charge state of given atom is smaller, it indicates ionic nature of bonding whereas the larger difference infers covalent bonding. When the 'X' atom in the PbX$_2$ structure is Iodine, the charge transfered from the Pb is less than 1e i.e., 0.93 e only. When the X move to more electronegative atom, there is an increase in charge depletion from Pb atom. Finally, in the PbF$_2$ monolayer, 1.53e transfered from Pb to fluorine atoms. Eventhough halide atoms have charge state -1e, the charge accumulated near halide atoms are less than 1e. iodine have less charge accumulation i.e., nearly 0.45 e. and the more electronegative fluorine have highest charge accumulation i.e., nearly 0.78 e. From the above analysis, it is found that the PbI$_2$ structure have more covalent nature. When the X atoms in the PbX$_2$ monolayer occupied with more electronegative atom than iodine, the covalent nature decreases and ionic nature increases. In PbX$_2$ monolayers, the charges accumulated at the top and bottom layers are equal amount due to the arrangement of same atoms in top and bottom layers.  But, due to the symmetrical charge distribution in top and bottom layers of the PbX$_2$ monolayers, there will not be any chance of dipole formation. \\

\begin{table*}
 \caption{Formation energy and Bader charge analysis of PbXY (X, Y = F, Cl, Br, I) Janus structures.}
\centering
   \begin{tabular}{ ||c c c c c c c  c c|| }
 \hline
  && \multicolumn{7}{ |c  ||}{\(\Delta\) Q}  \\[1ex]
 \hline
Structure & Formation Energy&  Pb (2+)& I (1-)& Br (1-)& Cl (1-)& F (1-) & \(\Delta\) C &\\[1ex] 
 \hline\hline
PbIBr & -2.66& 1.03 & -0.45 &-0.55 &&&0.10&\\[1ex]
PbICl &-2.91& 1.11 & -0.45& &-0.63 &&0.18&\\[1ex]
PbIF  &-4.54& 1.24 & -0.43 &&&- 0.79 &0.36&\\[1ex]
PbBrCl &-3.25& 1.21& & -0.56 & -0.63 &&0.07&\\[1ex]
PbBrF  &-4.94& 1.33 & &-0.53 && -0.79 &0.26&\\[1ex]
PbClF  & -5.21 & 1.41 & & & -0.62&-0.78 &0.16&\\[1ex] 
 \hline
 \end{tabular}  
\end{table*}
In case of Janus monolayers as per the data given in Table 2, we observed that the charge transfer from the Pb atoms to the halide atoms are not equal in top and bottom layers, due to the different electronegativities of halide atoms. But, in PbIF monolayer, Pb-F is more ionic and Pb-I is more covalent in nature. The inherent property of covalent and ionic character in the same structure may helps the materials to be used in different applications.
When we compare and analyse the differential charge density with the Bader charge results, we found that the completely spherical charge accumulation is shown by the ionic bonded atoms and the dumbbell shaped charge accumulation exhibited by the atoms which are connected by the covalent bond. The as discussed dumbbell shape in the charge accumulation is due to the spreading of charge between the atoms connected by covalent bond. This dispersed charge between the atoms act as charge transport path and improve the carrier mobility. The fast charge transfer supported by the covalent bonding will be beneficial to enhance the catalytic efficiency of the material.
For other Janus structures also, we  noticed that the charge accumulated at the top and  bottom layers have large difference, indicating the creation of large dipole moment from the more elecronegative atom to the less electronegative atom. This difference in accumulated charge near top and bottom layer of Janus structures are calculated using the equation given below ;
\begin{equation}
 \Large\Delta C = {C}_{top} - {C}_{bottom}
\end{equation}
 \(\Delta\)C  represents charge difference between top and bottom layer of Janus structures and quantify the electronegativity difference between different halide layers. 
 The \({C}_{top}\) and
 \({C}_{bottom}\) 
 represents charge accumulated near top and bottom layers of Janus structures.  Among the six Janus structures, this charge difference is high for PbIF and it is noted as 0.36 e. Similarly, the smallest charge difference between top and bottom layers of Janus structure is for PbBrCl and it is noted as 0.07 e. Due to the charge difference, there will be dipole formation, which will cause the electric field creation in the opposite direction of the created dipole. i.e., in the out of plane direction. This built-in electric field inside the Janus structures can successfully reduce the recombination of charge carriers. This will helps to enhance the lifetime of separated charge carriers which will improve the catalytic and photovoltaic capacity of the material.  Apart from this, the charge transfer direction we found from the Bader charge analysis is in accordance with the differential charge density plot.

\subsection{Energetics and the stability }
The energetics of Janus monolayers and their parent structures are studied with the help of formation energy. The formation energy of all these structures are calculated using the formula:\\
 \begin{equation}
\Large\Delta_{formation} = E_{monolayer} - \Sigma \mu N
 \end{equation}
Here, \(\Delta_{formation}\) and \(E_{monolayer}\)
represent formation energy and total energy of constructed monolayers\cite{42}.
Also, \(\mu\) represent chemical potential of  the different elements and N is the number of specified atoms in the unitcell.
For the chemical potential calculation, we have used Pb cubic  \((Fm\Bar{3}m) \)\cite{44},
I Orthorhombic (Immm)\cite{45}, Br Orthorhombic (Immm)\cite{46}, Cl Orthorhombic (Cmce)\cite{46b} and F monoclinic (C12/c1)\cite{46a} structures, having energy above the hull from material project database\cite{43}. The calculated chemical potential of each element is given in the Table S3.
 The calculated formation energy of PbX$_2$ monolayers are given in the table S2 and the formation energy of PbXY Janus structures are given in Table 2. For both PbX$_2$ monolayers as well as the newly designed Janus structures, the formation energy is negative indicating exothermic reaction and favorable energetics for the formation of these structures. Further, the formation energy of Janus structures are observed to be nearly average of their corresponding parent materials.\\
 \begin{figure*}
   \centering
\includegraphics[height= 18 cm]{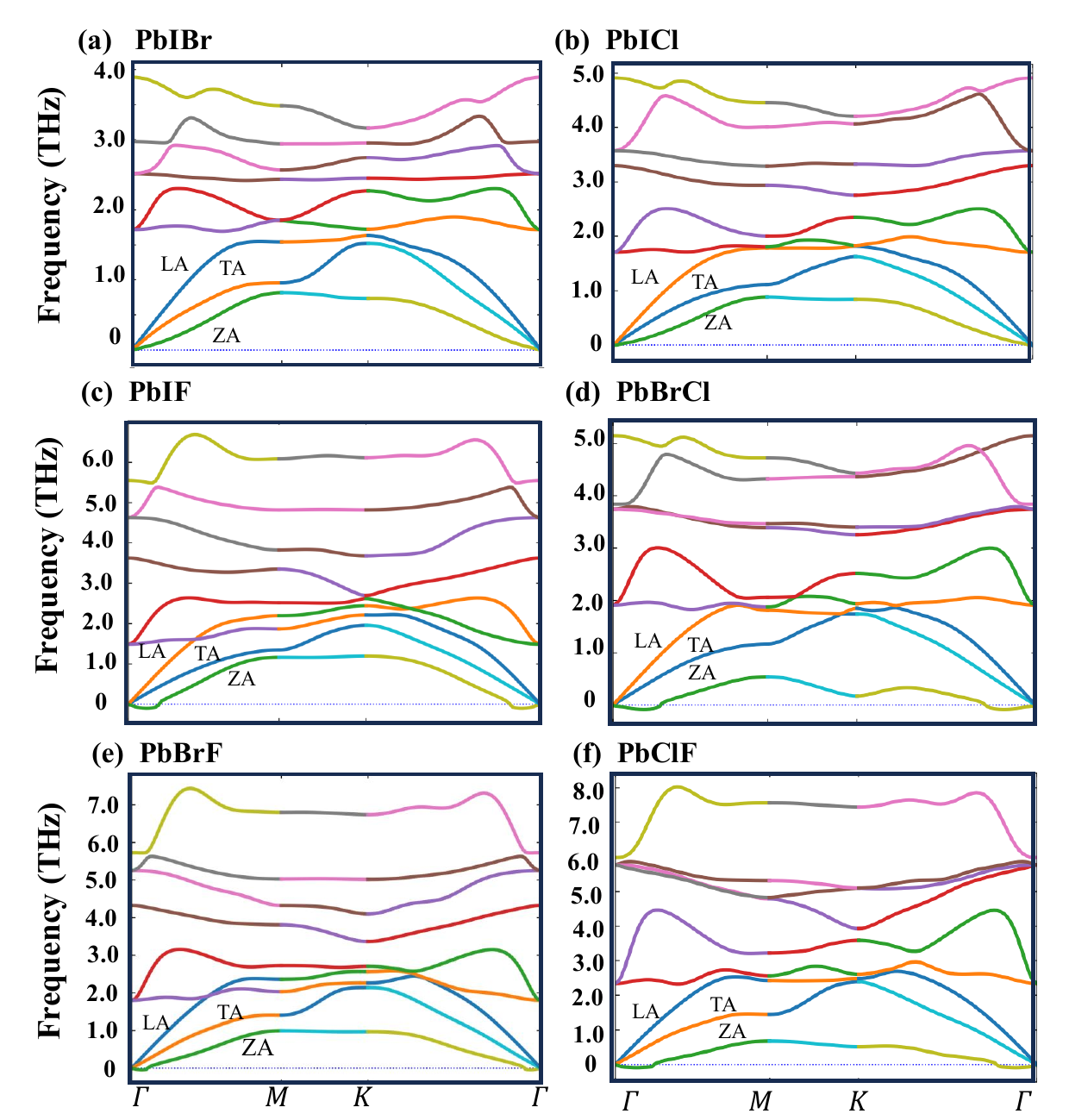}
  \caption{Phonon vibrational spectra of PbIBr (a) (6x6x1), PbICl (b) (5x5x1), PbIF (c) (7x7x1), PbBrCl (d) (7x7x1), PbBrF (e) (7x7x1) and PbClF (f) (7x7x1) monolayers plotted along the high symmetry points. The LA, TA and ZA modes represent longitudinal, transverse and out of plane acoustic modes. The supercell size of the phonon calculation given in bracket. Here, the negative frequencies indicates the imaginary mode.}
\end{figure*}
 Though different methods exist for the synthesis of 2D materials, the experimental synthesis of Janus structure is a great challenge infront of us. Among different methods to produce 2D materials, mechanical exfoliation method using scotch tape is well known, as it can produce very thin layered structure \cite{72}. When it comes to the Janus structures, the only reported method is controlled growth of material using chemical vapor deposition technique. 
The already reported MoSSe synthesis was from the MoS$_2$ monolayer by carefully removing one of the S layer and replacing with Se layer. In the same way, we can propose  the construction of new Janus monolayers from their parent structures.\\
Further, to understand the dynamical stability of the Janus structures, phonon dispersion curves are calculated along the path connecting  the high symmetry points in the Brillouin zone. We have performed detailed phonon study to elucidate the sensitive dependence of phonon frequencies with respect to supercell size. The supercell size we have considered are 3x3x1, 4x4x1, 5x5x1, 6x6x1, 7x7x1. We observed that for the smaller supercells like 3x3x1 and 4x4x1, Janus structures show significant amount of imaginary modes indicating instability of structures. But, by increasing the supercell size from 5x5x1, we could reduce the imaginary modes in the phonon spectra. The details of the phonon calculations like supercell size, kpoints used, error in the imaginary mode are given in Table S4. The calculated phonon spectra for different Janus structures are given in (Fig. S3, Fig. S4, Fig. S5, Fig. S6, Fig. S7, Fig. S8). As we discussed earlier, we noticed gradual decrease in imaginary modes with increase in supercell size. From the calculations performed, most stable phonon spectra are presented in Fig. 3. From these figures we found that the PbIBr (a)and PbICl (b)are showing no imaginary modes in 5x5x1 supercell and 6x6x1 supercell respectively. The absence of imaginary modes shows the dynamical stability of the structure. In the remaining four Janus structures, very less imaginary modes are observed. From our detailed comparative analysis of the supercell size and the error percentage, we show that in the larger supercell limit, these structures may show no imaginary modes\cite{50}. Moreover, the the reported errors for the most stable phonon structures are within the numerical error of the phonon calculation. \\
From the vibrational mode analysis, 6 optical and 3 acoustics modes are identified and analysed \cite{25}. The three acoustic phonon branches are inplane longitudinal acoustic mode (LA), transverse acoustic mode (TA) and out-of-plane acoustic mode (ZA)\cite{51}.  The linear dispersion of ZA phonon mode is an important feature of 2D material, which we noticed in highly stable 2D Janus structures of PbIBr and PbICl. For PbIBr Janus structure very small phonon gap is observed in between the acoustic phonon modes and optical phonon modes.
In contrast, no phonon gap is observed for other novel Janus structures. The phonon modes seems to be interlinked. This shows phonon-phonon interaction in theses Janus structures\cite{52}.  The highest frequency of optical phonon mode is varied for the different Janus structures. It is well known that, the maximum vibrational frequency is inversely propotional to the average of atomic mass of elements\cite{53}. As such, the highest maximum frequency vibrational phonon mode is observed in the PbClF Janus structure and the corresponding frequency is nearly 8 THz. Also, the lowest maximum frequency phonon mode is for PbIBr with corresponding frequency is nearly 4THz.  \\
 To understand the contribution of each atom of the Janus layers to the vibrational spectra, the atom resolved phonon DOS with most stable supercell is carried out (Fig. S9). From the analysis of phonon DOS, we could find that in PbIF Janus monolayer, the imaginary modes are contributed by Pb and I atoms. In case of PbBrCl, the small imaginary mode is shown by all the three atoms Pb, Br and Cl.
In contrast, in the other two Janus structures such as PbBrF and PbClF, the Pb atom only contributing to the unstable imaginary mode. Moreover, we found that at low frequency region the phonon modes corresponds to different atoms are same in PbIBr, PbICl and PbBrCl structures. Also, we noticed that the vibrational modes of elements having comparatively low weight, are observed to be at high frequency. 

\subsection{Electronic properties}
 The basic electronic and optical properties of a material are determined by the underlying electronic  structure. To analyse the electronic properties of PbXY Janus structures, electronic band structures are calculated using GGA functional. To get more accurate band structure, HSE calculations are performed. The GGA and HSE based electronic band structures are presented in Fig. 4. For the clarity, the conduction band minimum CBM positions of GGA and HSE band structure are marked with brown and yellow dotted lines. Also, the position of VBM marked by green dotted line. For comparison and to understand the changes in the electronic bandstructure, the GGA and HSE results parent structures are given in Fig. S10. When we compared the GGA and HSE band structure, it is noticed that the hybrid functional shifted the unoccupied bands to higher energy points. Thanks to this scissor correction, it could change the bandgap without altering the band dispersion considerably.  \\ 
 \begin{figure*}
 \centering
\includegraphics[height= 18cm]{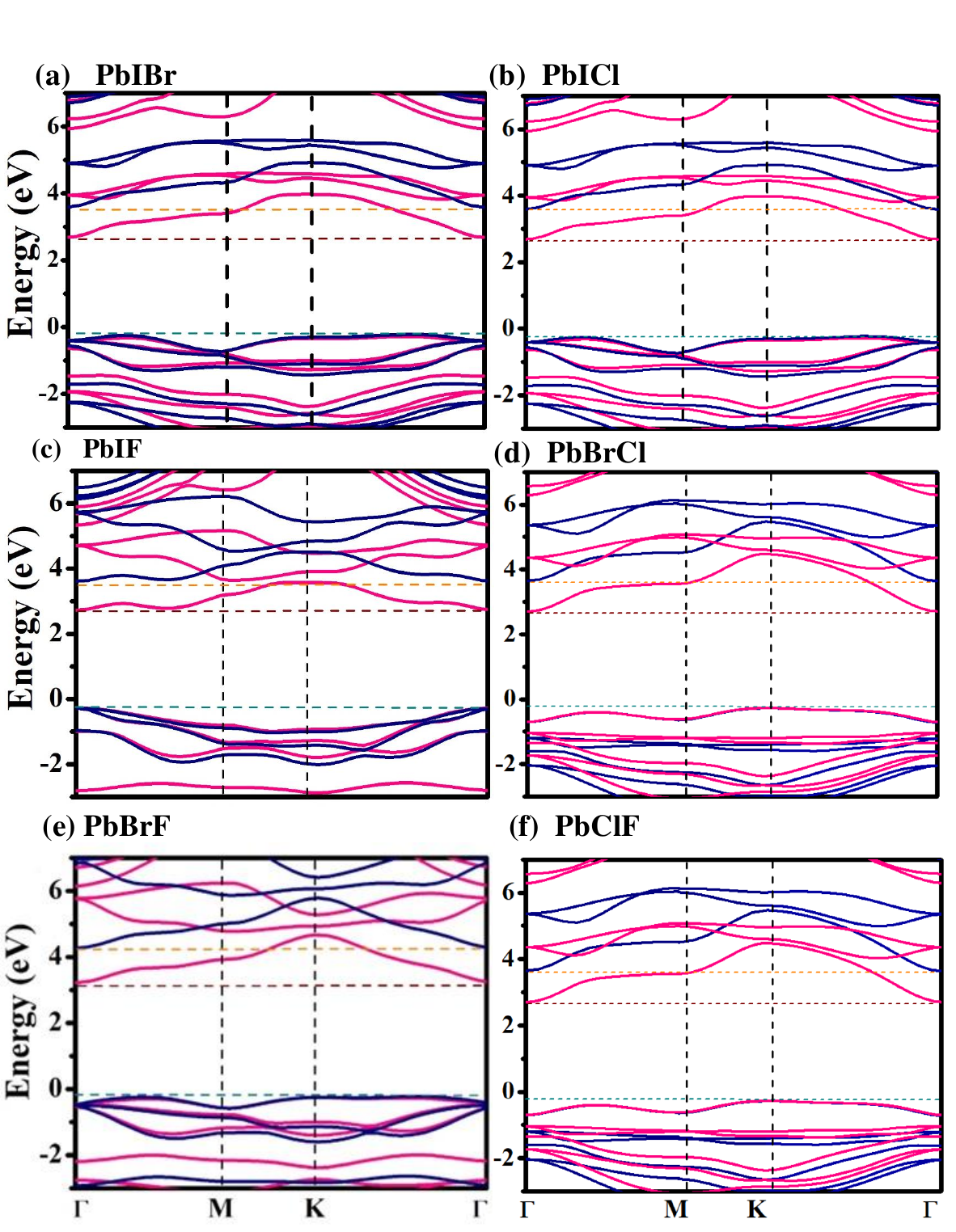}
  \caption{Basic electronic bandstructure of PbIBr(a), PbICl (b), PbIF (c), PbBrCl (d), PbBrF (e) and PbClF (f) calculated using GGA functional(pink line) and HSE functional (blue line). The blue dotted line represent VBM. The brown dotted line and yellow dotted line represents CBM calculated using GGA functional and HSE functional.} 
\end{figure*}
 In all the four parent structures, position of CBM is found to be at $\Gamma$ point and the valence band maximum at high symmetry point K, except for the PbI$_2$ monolayer. For PbI$_2$ monolayer, valence band maximum is in between $\Gamma$ and M point.
We have observed indirect band gap for all the monolayers. The measured bandgap corresponding to GGA and HSE functional are given in the Table S5.  The GGA band gap of the parent structures are in the range of 2.58 to 3.47 eV. The corrected bandgap using hybrid functional is in the range of 3.27 to 4.73 eV.\\
 From the electronic band structure analysis of Janus layers, it is found that all the Janus structures posess CBM position at $\Gamma$ point, which is similar to that of parent structures. The position of VBM is found to be in between $\Gamma$ and M for the PbIBr and PbICl monolayers. In PbIF, VBM is also observed to be at $\Gamma$ point. The VBM of remaining Janus structures are observed to be at K point. All the newly constructed Janus materials except PbIF are showing indirect band gap. In general, indirect bandgap materials are more prefered for the photocatalytic application, due to the the generated excitons take longer lifetime for recombination. In other words, indirect bandgap materials can reduce the recombination rate of charge carriers in effective way\cite{60}. By reducing the charge recombination rate, the generated charge carriers will get time to move to the surface of the photocatalyst and take part in water redox reaction. The calculated bandgap from the electronic band structure using GGA and HSE functional are presented in Table 3. Also the bandgap of parent materials are calculated and presented in Table S5. The GGA bandgaps of all the Janus structures are in the range of 2.75 to 3.56 eV. The given range will be beneficial for the absorption of visible solar spectrum, which also extended to UV range.  The HSE bandgap of the above mentioned Janus structures vary from 3.55 to 4.61 eV. The highest bandgap is measured for PbFCl and lowest bandgap is for the PbIBr monolayer. The conduction bands of the all Janus structures are highly dispersed compared to the valence bands, indicating the low effective mass of the electrons in conduction bands and the increased effective mass of holes in the valence band. Low effective mass of electrons in the conduction band is an indication of high electron mobility. Improved carrier mobility followed by the large spatial separation of electrons and holes due to the indirect band gap will also help for the improvement in the efficiency of the material as a photocatalyst and also as a material for the photovoltaic devices.\\
 
 \begin{table*}
    \caption{The basic electronic properties derived from band structure calculations ( Bandgap nature, Bandgap corresponding to GGA and HSE functional and effective mass of charge carriers) are presented. }
\centering
    \begin{tabular}{||cccccccc||}
    \hline
& \multicolumn{3}{c }{Bandgap (eV)} & \multicolumn{4}{c||}{Effective mass ($m^{*}$)} \\[1ex]
    \hline
    Structure & Bandgap nature & GGA& HSE& CBM-K & CBM-M & VBM-\(\Gamma\) & VBM-M \\[1ex]
    \hline\hline
PbIBr &Indirect&2.75&3.55& 0.305&0.305&1.53&1.41\\[1ex]
PbICl &Indirect&2.97&3.67& 0.339&0.340&2.56&1.56\\[1ex]
PbIF  &Direct&3.00&3.72& 0.491&0.491&1.69(VBM-K)&1.68\\[1ex]
PbBrCl &Indirect&2.97&3.89& 0.378&0.378&2.36&2.30\\[1ex]
PbBrF  &Indirect&3.48&4.35&0.527 &0.527&2.23&2.06\\[1ex]
PbClF  &Indirect&3.53&4.61& 0.678&0.677&1.93&1.92\\ [1ex]
    \hline
\end{tabular}
\end{table*}
For the quantitative analysis, the hole and electron  effective masses are calculated from VBM and CBM respectively. The calculated the effective mass of the Janus structures are given in Table 3 parent structures are reported in Table S5. For the Janus structures, the electron effective masses are in the range of 0.3 and 0.67 m$_e$. The less effective mass of electron indicate high carrier mobility. But the effective mass of holes in the VBM is calculated to be high and it is in between 1.5 and 2.3 m$_e$. The large difference between the effective mass of electrons and holes will lead to the  reduced recombination rates. This will also be helpful for the improvement in the catalytic activity of the material.
The partial density of states (PDOS) of all  parent structures (Fig. S11) and Janus structures are presented (Fig. S12). From the analysis of PDOS of both Janus and their parent structures, we found that valence band edge is mainly contributed by s and p orbital of Pb atom and p orbital of halide atoms. We have noticed the s-p hybridisation of both Pb and halide atoms in the valence band of all the structures. Also, the CBM is mainly contributed by p orbital of Pb atom in both Janus as well as their parent structures. Further, the orbital resolved p states (px, py, pz) is presented in Fig. 5 for the Janus structures and in Fig. S13 for the PbX$_2$ parent structures. In case of orbital resolved DOS of PbX$_2$ parent structures, we found the conduction band edge states of all the structures are build up of pz orbital.\\
\begin{figure*}
\centering
    \includegraphics[height= 20cm]{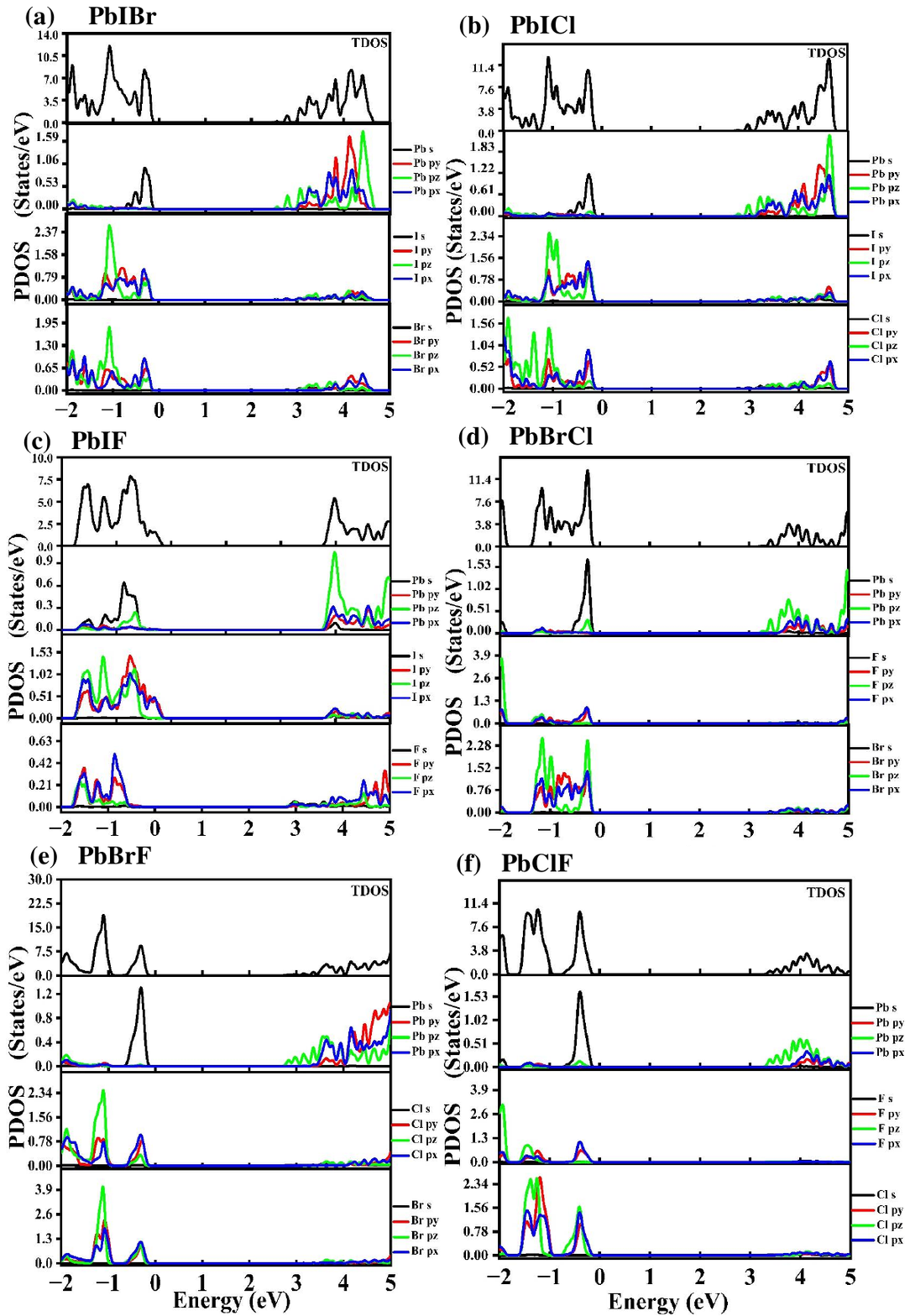}
    \caption{Partial density of states of PbIBr (a), PbICl (b), PbIF(c), PbBrCl(d), PbBrF(e), PbClBr (f) monolayers. The VBM is constructed out of s and p orbitals of Pb atom and p orbitals of halide atoms. The CBM is constructed out of pz orbital of Pb atom. The calculations are done using GGA functional.}
\end{figure*}
Further, from the analysis of orbital resolved DOS of Janus structures (Fig. 6), we found that the conduction band edge states are composed of pz orbital of Pb atom similar to that of PbX$_2$ monolayers. During optical excitation, the pz orbitals may help conduction electrons to move in the out of plane direction. The prominence of Janus structures compared to their parent structures, lies in the asymmetry in the structure. From the previous studies of differential charge density (Fig. 2) and the Bader charge analysis (Table 2), it is found that there is an asymmetry in charge distribution in the top and bottom layer of Janus structures. 
This difference in charge distribution will cause dipole formation and lead to electric field creation inside the Janus structures in the out-of plane direction. This optical excitation of charge carriers in the direction of built-in electric field can successfully increase the life time of separated charge carriers by mitigating the recombination rate. So that, photocatalytic efficiency of the constructed structures will be improved effectively. \\
\begin{figure*}
\includegraphics[  height= 6 cm]{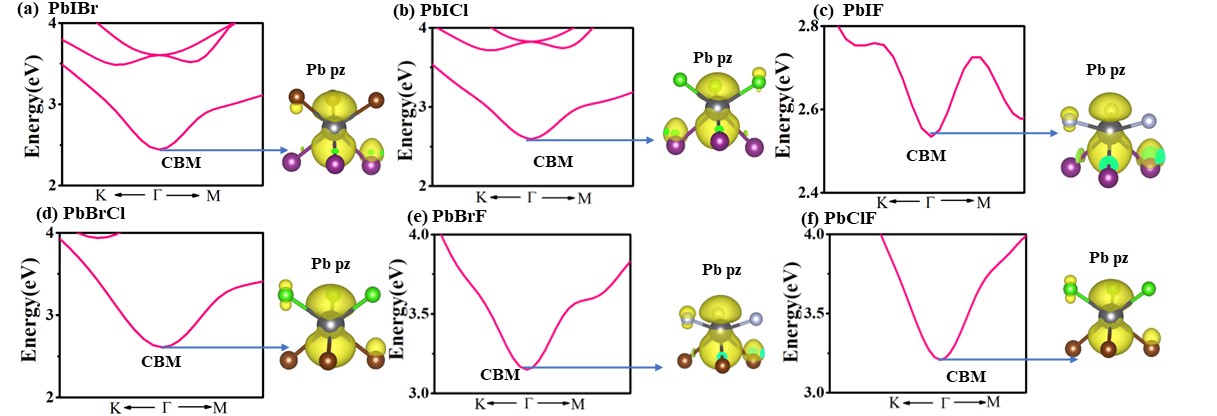}
\caption{Illustration of conduction band minimum and partial charge density representation at the CBM $(\Gamma)$ point of PbIBr (a), PbICl (b), PbIF (c), PbBrCl (d), PbBrF (e) and PbClF (f) monolayer with isosurface level 0.002 e/{\AA$^3$}. The partial charge density shows that the CBM is made of pz orbital of Pb atom. The calculations are performed using GGA functional.}
\end{figure*}
 To understand the orbital nature of the CBM, the partial charge density (PCD) is calculated at the gamma point. For the better clarity, the first unoccupied band and the PCD of all the Janus structures are illustrated in Fig. 6. Similarly, for the parent structures the above mentioned analysis is presented in Fig. S14. From the analysis of PCD of all structures, dumbbell shaped pz orbital near the Pb atom is observed at the CBM position of all the Janus structures as well as their parent structures. This illustrate CBM is composed of pz orbital confirming the orbital resolved DOS. 
 The necessary condition for the shift current generation is breaking of centrosymmetry. Since the Janus structures are naturally non centrosymmetric due to the asymmetry of the top and bottom layers, recently the Janus structures are being actively probed in the field of bulk photovoltaic effect using shift current. Other factors that supports shift current are covalent bonding which provides channels for electron transport and the favourable orbital character of conduction band edge state like presence of z oriented p orbital along the polarisation direction. From the Bader charge analysis, orbital resolved DOS and partial charge density at CBM, it is clear that the as designed Janus structures are suitable for the shift current generation and hence bulk photovoltaic effect.
 \subsection{Optical properties}
The optical absorbance study is performed for the PbXY Janus structures (Fig. 7) and their parent structures (Fig. S15) by calculating frequency dependent dielectric matrix with the help of high dense frequency grid points. From the calculated dielectric matrix, optical absorption coefficient is calculated from the imaginary and real part of the dielectric coefficients using the equation given below\cite{26}\cite{27}.
\\
\begin{equation}
    \centering
\Large I(\omega)=2\omega [\frac{[{\epsilon_{1}}^{2}(\omega)-{\epsilon_{2}}^2(\omega)]^{1/2}-\epsilon_{1}(\omega)}{2}]^{1/2}
\end{equation}
\\
Here, ${I(\omega)}$, ${\epsilon_{1}(\omega)},  {\epsilon}_{2}({\omega})$ are the optical absorption coefficient, real part of dielectric coefficient and the imaginary part of dielectric coefficient.
The average optical absorbance of PbX$_2$ monolayers along both x and y direction are calculated and presented in Fig. S15. 
From the analysis of the presented data, it is observed that the absorption peaks of PbI$_2$ is coming in the range of visible spectra. But, in the PbX$_2$ structures, when 'X' move towards more electronegative fluorine, the absorption peaks also shift towards high frequency range.\\

\begin{figure}
    \centering
\includegraphics[height= 9.75cm]{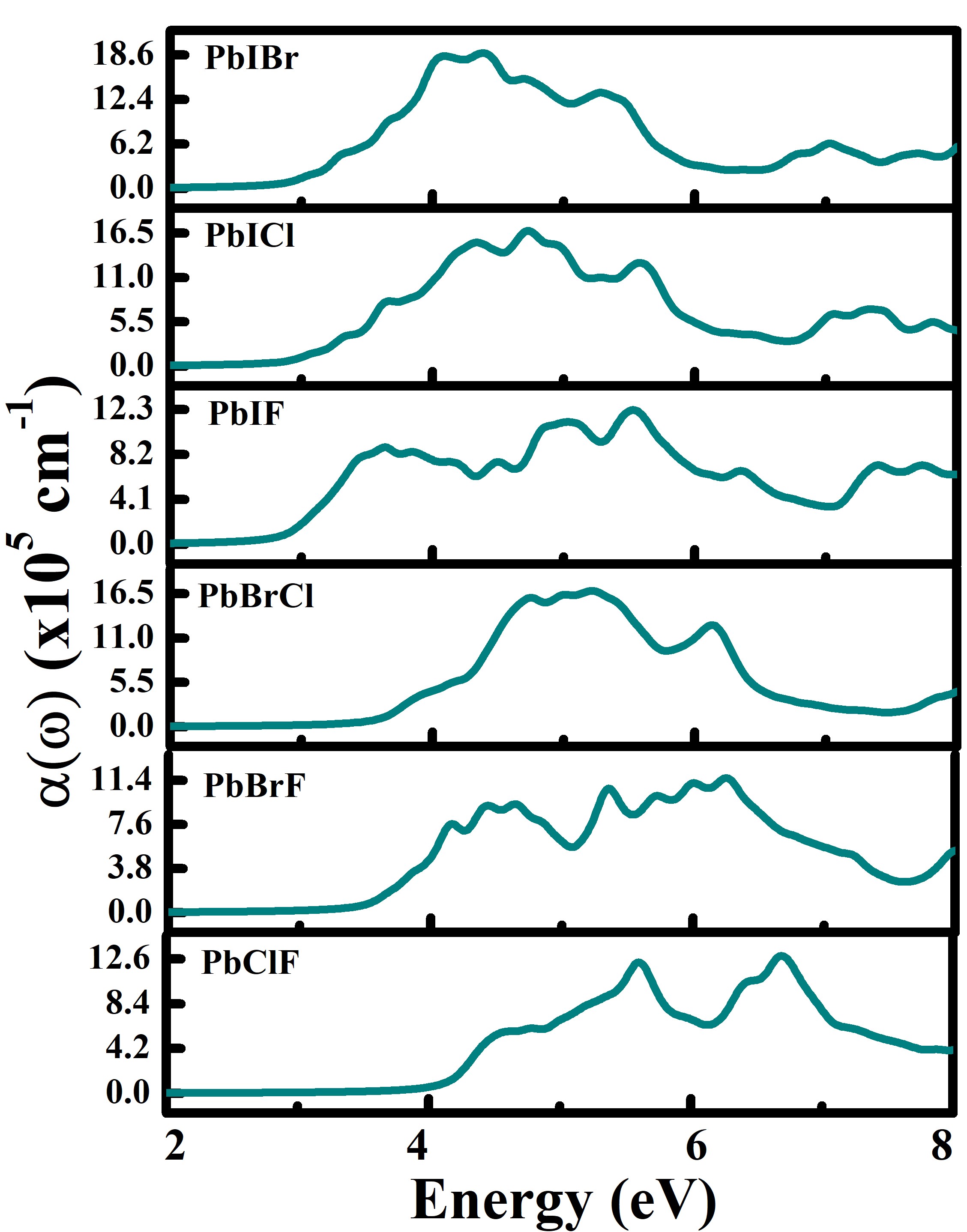}
  \caption{Comparison of Optical absorbance spectra of the designed PbXY Janus structures. Absorption peaks are observed at UV and visible range. The calculations are performed using GGA functional.}
\end{figure}
The optical absorbance ability of the Janus layers is presented in Fig. 7. It is observed that PbIBr, PbICl and PbIF shows comparitively good light absorption near-visible range of the solar spectra. For the PbBrCl, PbBrF and PbClF structures the absorption peaks are shifted to high energy range. So, the optical light absorption take place in the near visible range only. These structures can be considered as the near-visible light active photocatalyst. \\
 The light absorption can be further improved by applying strain or electric field. But, these methods are not investigated in this work. These are some of the already reported computational methods that can effectively tune the electronic and optical properties of monolayers\cite{59,60}. 

 \subsection{ Electrostatic potential and band alignment}
 
 The planar averaged electrostatic potential calculations are performed to analyse the alignment of fermi energy level, CBM and VBM with respect to the vacuum level.

  The calculations are performed by taking the average electrostatic potential along x and y direction and the electrostatic potential is calculated along z direction. The electrostatic potential diagram of PbX$_2$ parent structures (Fig. S16) and the Janus structures (Fig. S17) are  demonstrated.  The PbX$_2$ monolayers are symmetric in structure with similar top and bottom halide layers of Pb atom. So the electrostatic potential diagram also shown to be symmetric. But, there is a breaking of symmetry in the PbXY Janus structures with different halide layers at the top and bottom layers of Pb atom. So, the difference in the electrostatic potential of top and bottom layer can be clearly identified in the electrostatic potential diagram of Janus structures. This potential difference ($\Delta \phi$) is the reason for the the electric field inside the monolayer. This created potential difference is  high for the PbClF monolayers and small for the PbIBr and PbICl monolayers. 
  \\
 The water redox potentials with respect to vacuum levels are calculated using the equation given below\cite{82}:
 \begin{equation}
     \Large{E_{H^+/H_2}^{Red} = -4.44 +0.059 pH - {\Delta \phi}}
 \end{equation}
  \begin{equation}
     \Large{E_{O_2/H_2O}^{ox} = -5.67 +0.059 pH - {\Delta \phi}}
 \end{equation}
 
Here, \(E_{H^+/H_2}^{Red}\), \(E_{O_2/H_2O}^{ox}\)
and $\Delta \phi$
are the water reduction potential, oxidation potential and the electrostatic potential difference between the top and bottom layer respectively. At pH equal to zero, the water reduction and oxidation potentials are -4.44 eV and -5.67 eV respectively\cite{48}. These potentials are shifted by an amount $\Delta \phi$ in all the Janus structures. \\
To compare the band edge positions with respect to the above calculated water redox potentials, the conduction and the valence band edge positions were calculated with respect to vacuum level. The equations used to find the valence band and the conduction band edge positions are given below:\cite{24}. \\
 \begin{equation}
     \Large E_{C} = E_{BC} + (E_{g}/2)
 \end{equation}
  \begin{equation}
     \Large  E_{V} = E_{BC} - (E_{g}/2)
 \end{equation}
 \\
 Here, $E_{C},  E_{V} $
are the conduction band and valence band edge positions respectively. $E_{BC}$ is the band edge center, defined as negative of the work function. $E_{g}$ is the bandgap of the given material. Based on the electrostatic potential results as well as band structure (both GGA and HSE) calculations, band alignment of all PbX$_2$ structures are calculated and illustrated in Fig. S18. Also, the band alignment of PbXY Janus structures are also calculated and presented in Fig. 8. \cite{41}. In this figure, due to the created potential difference between the top and bottom layers of Janus structures, we introduce a shift in the water redox potential\cite{73}. This shift is equal to the potential difference between top and bottom layer. The CBM and VBM positions are calculated and given in the table S6. \\    
It was reported that for an ideal photocatalyst with good water splitting capacity, VBM and CBM should straddle the water oxidation and reduction potentials. As per this, we analysed the water redox capacity of this PbX$_2$ parent structures (Fig. S18) and PbXY Janus structures(Fig. 8).
 From the analysis of band alignment of PbX$_2$ structures, it is found that the VBM of all the structures straddle the water oxidation potential, hence it is suitable for the water oxidation reaction. In case of  CBM, the  band alignment calculations done with respect to GGA functional is not close to water reduction potential. But, after correcting the band alignment with HSE functional, the CBM of all the  structures are getting close to water reduction potential. This indicate that water reduction is possible in this material with the help of small external energy.  After correcting the CBM with HSE functional, the CBM of PbF$_2$  monolayer straddle the water reduction potential, indicating the structure is favourable for water reduction reaction. \\
  \begin{figure*}
\includegraphics[height=8cm]{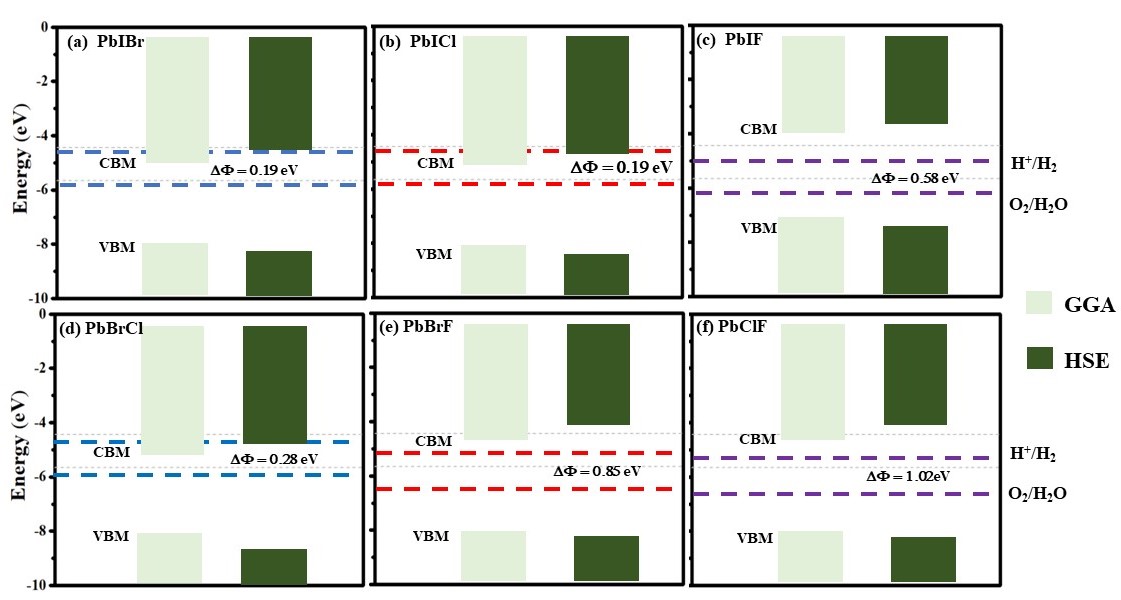}
  \caption{Band alignment of PbIBr (a), PbICl (b), PbIF (c), PbBrCl (d), PbBrF (e)and PbClF (f) monolayers with respect to vacuum level. The dotted lines represent water redox potentials at pH equal to zero. (water reduction at -4.44 eV, water oxidation at -5.67 eV). The water reduction potentials are shifted due to the potential difference($\Delta \Phi$) between the top and bottom layers. The shifted water redox potentials are presented with bright dashed lines.}
\end{figure*}
 From the analysis of Fig. 8, we can clearly understand that, VBM of all the Janus structures straddle the water oxidation potential. Also, CBM of all the Janus structures are close to the water reduction potential. For most of the Janus structures, even CBM straddles the water reduction potential which will be more favourable for water reduction reaction. The highly favourable band alignment achieved by the Janus structures is only due to the shift in water reduction potential introduced by the electrostatic potential difference between the top and bottom layers.  Here, it is clear that the potential difference in the Janus material have major role in helping the CBM and VBM of most of the Janus structures to straddle the water redox potential effectively. From the band alignment analysis, we can analyse that these structures are best for the water splitted hydrogen generation.

\section{Conclusion and Outlook}
In this project, we have successfully designed structurally and thermodynamically stable lead halide Janus structures PbXY (X, Y = F, Cl, Br, I) by the modification of PbX$_2$ monolayers. Some of the important points are summarized below: 

We found that the structural parameters such as lattice constants, bond lengths, bond angles and layer thickness show strong dependence on halide atoms present in both parent as well as Janus structures. In the Janus structures, the significant difference is observed between Pb-X and Pb-Y bond lengths. The two different sides of Janus structures may show very different behavior in the heterostructure arrangements giving rise to interesting new physics. Moreover, such a structural diversity can be exploited for the other applications as well.

The systematic study of dynamical stability of Janus structures are conducted with respect to different supercell sizes. The results indicate that atleast 5x5x1 supercell is necessary to show dynamical stability. Less than 5x5x1 supercells show significant imaginary modes. Two Janus structures PbIBr and PbICl show no imaginary modes whereas other four Janus structures show very little imaginary modes. The error analysis shows that these are within the numerical error limits of phonon calculations. Also, the systematic calculations reveal that the imaginary modes gradually decrease with increasing supercell size. The maximum supercell size considered is 7x7x1 due to constraints in the computational calculations. It is clear that in the limit of large supercell, these imaginary modes will disappear. 

The calculated formation energies are all negative indicating exothermic reaction and favorable energetics for the formation of these compounds. The experimental condition at which the Janus structures form can be obtained by calculating formation energy as a function of chemical potentials of constituent atoms.  


The bonding nature and charge transfer mechanism studied using differential charge density calculations and Bader charge analysis shows that parent as well as Janus structures show covalent nature of bonding except in case of lighter and more electronegative atoms like F and Cl. Strikingly, the Janus structure show very strong dual nature in bonding, one side ionic and other side covalent, when electronegativity difference between top and bottom layers are large. The difference in electronegativity will cause dipole formation along the out of plane direction and helps to separate the charge carriers generated upon solar light absorption. This will reduce the recombination rate. Larger the dipole, smaller the recombination rate of charge carriers. 

The GGA and HSE levels of electronic bandstructure calculations indicate that Janus structures show both direct and indirect nature of bandgap. Large tunability of band gap from 2.5 to 3.5 eV is observed with GGA calculations. HSE calculations predict wider band gaps. The charge carrier mobilities can be inferred from the effective masses. The smaller the effective masses larger the mobility. Electron effective masses are low and in the range of ideally suitable for photovoltaic and photocatalysis. The hole effective masses are found to be higher and the large difference between the electron and hole effective masses can easily separate the charge carrier and hence reduce the recombination rates. 

The PDOS shows that the valence band edge is composed of p orbitals of both Pb and halide atoms. The conduction band edge states are composed of Pb p orbitals. Further the orbital resolved DOS shows that the conduction band edge is composed pz orbital of Pb atoms. In order to understand the exact orbital character of CBM, partial charge densities are calculated at CBM. This again confirms the pz orbital of Pb atoms at CBM. This clearly shows that upon light absorption, the charge carriers can be swiftly separated in the out of plane direction due to the presence pz orbital at the conduction band edge states coinciding with the polarization direction. Further, the covalent nature also helps charge carrier transport along the out of plane directions. All these properties may also help bulk photovoltaic effect (BPVE) through shift current generation in the material as Janus structures show breaking of centrosymmetry. 

The frequency dependent optical absorption calculations show that Janus structures are suitable for visible light absorption as well as UV light. 

Finally, to understand the water splitting capacity of parent as well Janus structures, band alignments are calculated with respect to vacuum level. The obtained CBM and VBM are compared with water redox potentials. This is mainly due to the electric potential difference between the top and bottom layer. In contrast the parent PbI$_2$ monolayers are less or not appropriate for water splitting. We find that all the Janus structures are suitable for the water splitting redox reactions as both VBM and CBM straddle oxidation and reduction potentials. \\
As of now, shift current is discussed only with bulk photovoltaic effect. Here we show that these materials can also support photocatalysis.

For the future research, we propose the designed Janus structures for the other energy applications such as thermoelectrics, energy storage and for the multifunctional devices. The properties of the material can be further fine tuned by applying strain or electric field. Finally, we would also like to propose design of new Janus halide structures with non toxic nature, instead of Pb based systems. The theoretical studies we conducted on the the novel prototypical Janus structures PbXY will be helpful to understand different inherent properties and physics of the new structures.


\begin{acknowledgement}
Dr. D. Murali is thankful to the DST-SERB for financial support (CRG/2022/006778, 
MTR/2020/000551).

\end{acknowledgement}

\begin{suppinfo}
The additional information such as phonon DOS and electrostaic potential energy diagrams of designed Janus structures are given in the supplementary information.
Also, the structural, electronic and optical properties of all the parent structures are given in detail.
\end{suppinfo}

\bibliography{achemso-demo}

\providecommand{\latin}[1]{#1}
\makeatletter
\providecommand{\doi}
  {\begingroup\let\do\@makeother\dospecials
  \catcode`\{=1 \catcode`\}=2 \doi@aux}
\providecommand{\doi@aux}[1]{\endgroup\texttt{#1}}
\makeatother
\providecommand*\mcitethebibliography{\thebibliography}
\csname @ifundefined\endcsname{endmcitethebibliography}  {\let\endmcitethebibliography\endthebibliography}{}
\begin{mcitethebibliography}{80}
\providecommand*\natexlab[1]{#1}
\providecommand*\mciteSetBstSublistMode[1]{}
\providecommand*\mciteSetBstMaxWidthForm[2]{}
\providecommand*\mciteBstWouldAddEndPuncttrue
  {\def\EndOfBibitem{\unskip.}}
\providecommand*\mciteBstWouldAddEndPunctfalse
  {\let\EndOfBibitem\relax}
\providecommand*\mciteSetBstMidEndSepPunct[3]{}
\providecommand*\mciteSetBstSublistLabelBeginEnd[3]{}
\providecommand*\EndOfBibitem{}
\mciteSetBstSublistMode{f}
\mciteSetBstMaxWidthForm{subitem}{(\alph{mcitesubitemcount})}
\mciteSetBstSublistLabelBeginEnd
  {\mcitemaxwidthsubitemform\space}
  {\relax}
  {\relax}

\bibitem[de~Gennes~PG.(1992)]{1_1}
de~Gennes~PG. soft matter. \emph{Soft matter. Science.} \textbf{1992}, \emph{256}, 495--497\relax
\mciteBstWouldAddEndPuncttrue
\mciteSetBstMidEndSepPunct{\mcitedefaultmidpunct}
{\mcitedefaultendpunct}{\mcitedefaultseppunct}\relax
\EndOfBibitem
\bibitem[Andreas~Walther(2013)]{2}
Andreas~Walther,~A. H. E.~M. Janus Particles: Synthesis, Self-Assembly, Physical Properties, and Applications. \emph{Chem. Rev.} \textbf{2013}, \emph{113}, 5194--5261\relax
\mciteBstWouldAddEndPuncttrue
\mciteSetBstMidEndSepPunct{\mcitedefaultmidpunct}
{\mcitedefaultendpunct}{\mcitedefaultseppunct}\relax
\EndOfBibitem
\bibitem[Montes-García and Samorì(2022)Montes-García, and Samorì]{29}
Montes-García,~V.; Samorì,~P. Janus 2D materials via asymmetric molecular functionalization. \emph{Chem. Sci.} \textbf{2022}, \emph{13}, 315--328\relax
\mciteBstWouldAddEndPuncttrue
\mciteSetBstMidEndSepPunct{\mcitedefaultmidpunct}
{\mcitedefaultendpunct}{\mcitedefaultseppunct}\relax
\EndOfBibitem
\bibitem[Lei \latin{et~al.}(2020)Lei, Zhenjingfeng, Tian, Ruikun, Huide, Zhinan, Han, and Xiao]{30}
Lei,~Z.; Zhenjingfeng,~Y.; Tian,~G.; Ruikun,~P.; Huide,~W.; Zhinan,~G.; Han,~Z.; Xiao,~F. Recent advances in emerging Janus two-dimensional materials: from fundamental physics to device applications. \emph{J. Mater. Chem. A} \textbf{2020}, \emph{8}, 8813--8830\relax
\mciteBstWouldAddEndPuncttrue
\mciteSetBstMidEndSepPunct{\mcitedefaultmidpunct}
{\mcitedefaultendpunct}{\mcitedefaultseppunct}\relax
\EndOfBibitem
\bibitem[Georgakilas \latin{et~al.}(2012)Georgakilas, Otyepka, Bourlinos, Chandra, Kim, Kemp, Hobza, Zboril, and Kim]{4_1}
Georgakilas,~V.; Otyepka,~M.; Bourlinos,~A.; Chandra,~V.; Kim,~N.; Kemp,~K.; Hobza,~P.; Zboril,~R.; Kim,~K. Functionalization of Graphene: Covalent and Non-Covalent Approaches, Derivatives and Applications. \emph{Chemical reviews} \textbf{2012}, \emph{112}\relax
\mciteBstWouldAddEndPuncttrue
\mciteSetBstMidEndSepPunct{\mcitedefaultmidpunct}
{\mcitedefaultendpunct}{\mcitedefaultseppunct}\relax
\EndOfBibitem
\bibitem[Zhou \latin{et~al.}(2009)Zhou, Wang, Sun, Chen, Kawazoe, and Jena]{4_2}
Zhou,~J.; Wang,~Q.; Sun,~Q.; Chen,~X.; Kawazoe,~Y.; Jena,~P. Ferromagnetism in Semihydrogenated Graphene Sheet. \emph{Nano letters} \textbf{2009}, \emph{9}, 3867--70\relax
\mciteBstWouldAddEndPuncttrue
\mciteSetBstMidEndSepPunct{\mcitedefaultmidpunct}
{\mcitedefaultendpunct}{\mcitedefaultseppunct}\relax
\EndOfBibitem
\bibitem[Kim \latin{et~al.}(2014)Kim, Noor-A-Alam, Son, and Shin]{4_3}
Kim,~H.~J.; Noor-A-Alam,~M.; Son,~J.~Y.; Shin,~Y.-H. Origin of piezoelectricity in monolayer halogenated graphane piezoelectrics. \emph{Chemical Physics Letters} \textbf{2014}, \emph{603}, 62--66\relax
\mciteBstWouldAddEndPuncttrue
\mciteSetBstMidEndSepPunct{\mcitedefaultmidpunct}
{\mcitedefaultendpunct}{\mcitedefaultseppunct}\relax
\EndOfBibitem
\bibitem[Yujin \latin{et~al.}(2018)Yujin, Mingye, Haiping, Tingjun, Lu, Youyong, and Shuit-Tong]{4}
Yujin,~J.; Mingye,~Y.; Haiping,~L.; Tingjun,~H.; Lu,~W.; Youyong,~L.; Shuit-Tong,~L. Janus Structures of Transition Metal Dichalcogenides as the Heterojunction Photocatalysts for Water Splitting. \emph{J. Phys. Chem. C} \textbf{2018}, \emph{122}, 3123--3129\relax
\mciteBstWouldAddEndPuncttrue
\mciteSetBstMidEndSepPunct{\mcitedefaultmidpunct}
{\mcitedefaultendpunct}{\mcitedefaultseppunct}\relax
\EndOfBibitem
\bibitem[Khadijatul \latin{et~al.}(2022)Khadijatul, Rafiqul, Sakib, Shaffatul, and Tanvir]{5}
Khadijatul,~K.; Rafiqul,~I.~M.; Sakib,~H. K.~M.; Shaffatul,~I.~M.; Tanvir,~H.~M. Study of Two-Dimensional Janus {WXY} ({X,Y}= {S, S}e, and {T}e) Trilayer Homostructures for Photovoltaic Applications Using DFT Screening of Different Stacking Patterns. \emph{ACS Omega} \textbf{2022}, \emph{7}, 12947-- 12955\relax
\mciteBstWouldAddEndPuncttrue
\mciteSetBstMidEndSepPunct{\mcitedefaultmidpunct}
{\mcitedefaultendpunct}{\mcitedefaultseppunct}\relax
\EndOfBibitem
\bibitem[Bikerouin and Balli"(2022)Bikerouin, and Balli"]{32}
Bikerouin,~M.; Balli",~M. Janus transition-metal dichalcogenides heterostructures for highly efficient excitonic solar cells. \emph{Applied Surface Science} \textbf{2022}, \emph{598}, 153835\relax
\mciteBstWouldAddEndPuncttrue
\mciteSetBstMidEndSepPunct{\mcitedefaultmidpunct}
{\mcitedefaultendpunct}{\mcitedefaultseppunct}\relax
\EndOfBibitem
\bibitem[Ju \latin{et~al.}(2020)Ju, Bie, Tang, Shang, and Kou]{55}
Ju,~L.; Bie,~M.; Tang,~X.; Shang,~J.; Kou,~L. Janus WSSe Monolayer: An Excellent Photocatalyst for Overall Water Splitting. \emph{ACS Applied Materials \& Interfaces} \textbf{2020}, \emph{12}, 29335--29343, PMID: 32519846\relax
\mciteBstWouldAddEndPuncttrue
\mciteSetBstMidEndSepPunct{\mcitedefaultmidpunct}
{\mcitedefaultendpunct}{\mcitedefaultseppunct}\relax
\EndOfBibitem
\bibitem[Cui \latin{et~al.}(2022)Cui, Yang, Peng, and Zhang]{6}
Cui,~H.; Yang,~T.; Peng,~X.; Zhang,~G. First-principles screening upon Janus {P}t{XY} ({X, Y = S, S}e and {T}e) monolayer under applied biaxial strains and electric fields. \emph{Journal of Materials Research and Technology} \textbf{2022}, \emph{18}, 1218--1229\relax
\mciteBstWouldAddEndPuncttrue
\mciteSetBstMidEndSepPunct{\mcitedefaultmidpunct}
{\mcitedefaultendpunct}{\mcitedefaultseppunct}\relax
\EndOfBibitem
\bibitem[Xun \latin{et~al.}(2023)Xun, Xiaohao, Deyan, and Xiaoshuang]{33}
Xun,~G.; Xiaohao,~Z.; Deyan,~S.; Xiaoshuang,~C. First-Principles Study of Structural and Electronic Properties of Monolayer {P}t{X}$_2$ and Janus {P}t{XY} ({X, Y = S, Se, and Te}) via Strain Engineering. \emph{ACS Omega} \textbf{2023}, \emph{8}, 5715--5721\relax
\mciteBstWouldAddEndPuncttrue
\mciteSetBstMidEndSepPunct{\mcitedefaultmidpunct}
{\mcitedefaultendpunct}{\mcitedefaultseppunct}\relax
\EndOfBibitem
\bibitem[Rui \latin{et~al.}(2019)Rui, Yandong, Baibiao, and Ying]{54}
Rui,~P.; Yandong,~M.; Baibiao,~H.; Ying,~D. Two-dimensional Janus PtSSe for photocatalytic water splitting under the visible or infrared light. \emph{J. Mater. Chem. A} \textbf{2019}, \emph{7}, 603--610\relax
\mciteBstWouldAddEndPuncttrue
\mciteSetBstMidEndSepPunct{\mcitedefaultmidpunct}
{\mcitedefaultendpunct}{\mcitedefaultseppunct}\relax
\EndOfBibitem
\bibitem[Meiqiu \latin{et~al.}(2023)Meiqiu, Yi, Xingao, Yun, and Xuhai]{6_1}
Meiqiu,~X.; Yi,~S.; Xingao,~L.; Yun,~D.; Xuhai,~L. An Ab Initio Study of Two Dimensional SnX$_2$ and Janus SnXY (X = S, Se) Nanosheets as Potential Photocatalysts for Water Splitting. \emph{ACS Appl. Nano Mater.} \textbf{2023}, \emph{6}, 10569--10580\relax
\mciteBstWouldAddEndPuncttrue
\mciteSetBstMidEndSepPunct{\mcitedefaultmidpunct}
{\mcitedefaultendpunct}{\mcitedefaultseppunct}\relax
\EndOfBibitem
\bibitem[Pan \latin{et~al.}(2020)Pan, Yixin, Hao, Hongyu, Hui-Xiong, Zhongming, Hai-Bin, and Jian-Bai]{34}
Pan,~W.; Yixin,~Z.; Hao,~L.; Hongyu,~W.; Hui-Xiong,~D.; Zhongming,~W.; Hai-Bin,~W.; Jian-Bai,~X. Quasiparticle Band Structure and Optical Properties of the Janus Monolayer and Bilayer {S}n{SS}e. \emph{The Journal of Physical Chemistry C} \textbf{2020}, \emph{124}, 23832-- 23838\relax
\mciteBstWouldAddEndPuncttrue
\mciteSetBstMidEndSepPunct{\mcitedefaultmidpunct}
{\mcitedefaultendpunct}{\mcitedefaultseppunct}\relax
\EndOfBibitem
\bibitem[Dat and Vu(2022)Dat, and Vu]{56}
Dat,~V.~D.; Vu,~T.~V. Janus monolayer HfSO with improved optical properties as a novel material for photovoltaic and photocatalyst applications. \emph{New J. Chem.} \textbf{2022}, \emph{46}, 1557--1568\relax
\mciteBstWouldAddEndPuncttrue
\mciteSetBstMidEndSepPunct{\mcitedefaultmidpunct}
{\mcitedefaultendpunct}{\mcitedefaultseppunct}\relax
\EndOfBibitem
\bibitem[Zhang \latin{et~al.}(2021)Zhang, Qiu, Guo, Wu, Zhu, Zheng, Li, Wang, Chen, and Yu]{57}
Zhang,~F.; Qiu,~J.; Guo,~H.; Wu,~L.; Zhu,~B.; Zheng,~K.; Li,~H.; Wang,~Z.; Chen,~X.; Yu,~J. Theoretical investigations of novel Janus Pb2SSe monolayer as a potential multifunctional material for piezoelectric{,} photovoltaic{,} and thermoelectric applications. \emph{Nanoscale} \textbf{2021}, \emph{13}, 15611--15623\relax
\mciteBstWouldAddEndPuncttrue
\mciteSetBstMidEndSepPunct{\mcitedefaultmidpunct}
{\mcitedefaultendpunct}{\mcitedefaultseppunct}\relax
\EndOfBibitem
\bibitem[Xu \latin{et~al.}(2020)Xu, Yao, Yin, Cao, Chen, and Wei]{7_1}
Xu,~Y.; Yao,~Y.; Yin,~W.-J.; Cao,~J.; Chen,~M.; Wei,~X. Intrinsic defect engineered Janus MoSSe sheet as a promising photocatalyst for water splitting. \emph{RSC Advances} \textbf{2020}, \emph{10}, 10816--10825\relax
\mciteBstWouldAddEndPuncttrue
\mciteSetBstMidEndSepPunct{\mcitedefaultmidpunct}
{\mcitedefaultendpunct}{\mcitedefaultseppunct}\relax
\EndOfBibitem
\bibitem[Bikerouin \latin{et~al.}(2023)Bikerouin, Chdil, and Balli]{7_2}
Bikerouin,~M.; Chdil,~O.; Balli,~M. Solar cells based on 2D Janus group-III chalcogenide van der Waals heterostructures. \emph{Nanoscale} \textbf{2023}, \emph{15}, 7126--7138\relax
\mciteBstWouldAddEndPuncttrue
\mciteSetBstMidEndSepPunct{\mcitedefaultmidpunct}
{\mcitedefaultendpunct}{\mcitedefaultseppunct}\relax
\EndOfBibitem
\bibitem[Patel \latin{et~al.}(2020)Patel, Singh, Sonvane, Thakor, and Ahuja]{58}
Patel,~A.; Singh,~D.; Sonvane,~Y.; Thakor,~P.~B.; Ahuja,~R. High Thermoelectric Performance in Two-Dimensional Janus Monolayer Material WS-X (X = Se and Te). \emph{ACS Applied Materials \& Interfaces} \textbf{2020}, \emph{12}, 46212--46219, PMID: 32931245\relax
\mciteBstWouldAddEndPuncttrue
\mciteSetBstMidEndSepPunct{\mcitedefaultmidpunct}
{\mcitedefaultendpunct}{\mcitedefaultseppunct}\relax
\EndOfBibitem
\bibitem[Mukesh \latin{et~al.}(2023)Mukesh, Raman, and Ashok]{9}
Mukesh,~J.; Raman,~S.; Ashok,~K. Janus \textbeta-{P}d{XY} ({X,Y} = {S, Se, Te}) materials with high anisotropic thermoelectric performance. \emph{Nanoscale} \textbf{2023}, \emph{15}, 5964--5975\relax
\mciteBstWouldAddEndPuncttrue
\mciteSetBstMidEndSepPunct{\mcitedefaultmidpunct}
{\mcitedefaultendpunct}{\mcitedefaultseppunct}\relax
\EndOfBibitem
\bibitem[Mehta \latin{et~al.}(2023)Mehta, Modi, Khengar, Jariwala, and Thakor]{8}
Mehta,~D.; Modi,~N.; Khengar,~S.; Jariwala,~P.; Thakor,~P. A {G}a{A}l{S}$_2$ Janus Monolayer as a Promising Candidate for Optoelectronic Devices. \emph{Materials Today: Proceedings} \textbf{2023}, \relax
\mciteBstWouldAddEndPunctfalse
\mciteSetBstMidEndSepPunct{\mcitedefaultmidpunct}
{}{\mcitedefaultseppunct}\relax
\EndOfBibitem
\bibitem[Zhang \latin{et~al.}(2022)Zhang, Xia, Li, Li, Fu, Cheng, and Pan]{37}
Zhang,~L.; Xia,~Y.; Li,~X.; Li,~L.; Fu,~X.; Cheng,~J.; Pan,~R. {Janus two-dimensional transition metal dichalcogenides}. \emph{Journal of Applied Physics} \textbf{2022}, \emph{131}, 230902\relax
\mciteBstWouldAddEndPuncttrue
\mciteSetBstMidEndSepPunct{\mcitedefaultmidpunct}
{\mcitedefaultendpunct}{\mcitedefaultseppunct}\relax
\EndOfBibitem
\bibitem[Zhenxing \latin{et~al.}(2017)Zhenxing, Xiaobin, Zhongjun, Yuyan, Lu, and Lei]{9_1}
Zhenxing,~W.; Xiaobin,~Y.; Zhongjun,~C.; Yuyan,~L.; Lu,~S.; Lei,~J. Simply realizing “water diode” Janus membranes for multifunctional smart applications. \emph{Mater. Horiz.} \textbf{2017}, \emph{4}, 701--708\relax
\mciteBstWouldAddEndPuncttrue
\mciteSetBstMidEndSepPunct{\mcitedefaultmidpunct}
{\mcitedefaultendpunct}{\mcitedefaultseppunct}\relax
\EndOfBibitem
\bibitem[Hou \latin{et~al.}(2022)Hou, Xue, Qiu, Wang, and Wu]{11}
Hou,~Y.; Xue,~F.; Qiu,~L.; Wang,~Z.; Wu,~R. Multifunctional two-dimensional van der Waals Janus magnet Cr-based dichalcogenide halides. \emph{npj Computational Materials} \textbf{2022}, \emph{8}, 120\relax
\mciteBstWouldAddEndPuncttrue
\mciteSetBstMidEndSepPunct{\mcitedefaultmidpunct}
{\mcitedefaultendpunct}{\mcitedefaultseppunct}\relax
\EndOfBibitem
\bibitem[Sun \latin{et~al.}(2022)Sun, Xinru~Li, Wei, Guo, and Wang]{11a}
Sun,~Z.; Xinru~Li,~J.~L.; Wei,~Y.; Guo,~H.; Wang,~J. Intrinsic bitunable magnetism/polarity behavior in 2D Janus Cr$_2$I$_3$Y3 (Y = F, Cl, or Br) systems. \emph{npj 2D Mater Appl.} \textbf{2022}, \emph{6}, 69\relax
\mciteBstWouldAddEndPuncttrue
\mciteSetBstMidEndSepPunct{\mcitedefaultmidpunct}
{\mcitedefaultendpunct}{\mcitedefaultseppunct}\relax
\EndOfBibitem
\bibitem[Tayran \latin{et~al.}(2023)Tayran, Mogulkoc, and Çakmak]{12}
Tayran,~C.; Mogulkoc,~Y.; Çakmak,~M. Novel two-dimensional Janus {YMN} ({M= I, Br} and {N= Cl, Br}) monolayers. \emph{Materials Science in Semiconductor Processing} \textbf{2023}, \emph{166}, 107745\relax
\mciteBstWouldAddEndPuncttrue
\mciteSetBstMidEndSepPunct{\mcitedefaultmidpunct}
{\mcitedefaultendpunct}{\mcitedefaultseppunct}\relax
\EndOfBibitem
\bibitem[Peng \latin{et~al.}(2019)Peng, Mei, Zhang, Shao, Xu, Ni, Jin, Soukoulis, and Zhu]{70}
Peng,~B.; Mei,~H.; Zhang,~H.; Shao,~H.; Xu,~K.; Ni,~G.; Jin,~Q.; Soukoulis,~C.~M.; Zhu,~H. High thermoelectric efficiency in monolayer PbI$_2$ from 300 K to 900 K. \emph{Inorg. Chem. Front.} \textbf{2019}, \emph{6}, 920--928\relax
\mciteBstWouldAddEndPuncttrue
\mciteSetBstMidEndSepPunct{\mcitedefaultmidpunct}
{\mcitedefaultendpunct}{\mcitedefaultseppunct}\relax
\EndOfBibitem
\bibitem[E. \latin{et~al.}(2023)E., Pooja, Assa, D., and Murali]{68}
E.,~S.~A.; Pooja,~V.; Assa,~A. S.~D.; D.,~A.~R.; Murali,~D. Effective tailoring of the MoS$_2$ layer number on the surface of CdS nanorods for boosting hydrogen production rate. \emph{Dalton Trans.} \textbf{2023}, \emph{52}, 5297--5311\relax
\mciteBstWouldAddEndPuncttrue
\mciteSetBstMidEndSepPunct{\mcitedefaultmidpunct}
{\mcitedefaultendpunct}{\mcitedefaultseppunct}\relax
\EndOfBibitem
\bibitem[Pooja \latin{et~al.}(2022)Pooja, E., Assa, D., and Amaranatha~Reddy]{69}
Pooja,~V.; E.,~S.~A.; Assa,~A. S.~D.; D.,~M.; Amaranatha~Reddy,~D. Regulating the charge carrier transport rate via bridging ternary heterojunctions to enable CdS nanorods’ solar-driven hydrogen evolution. \emph{Dalton Trans.} \textbf{2022}, \emph{51}, 18693--18707\relax
\mciteBstWouldAddEndPuncttrue
\mciteSetBstMidEndSepPunct{\mcitedefaultmidpunct}
{\mcitedefaultendpunct}{\mcitedefaultseppunct}\relax
\EndOfBibitem
\bibitem[Ju \latin{et~al.}(2020)Ju, Bie, Shang, Tang, and Kou]{64}
Ju,~L.; Bie,~M.; Shang,~J.; Tang,~X.; Kou,~L. Janus transition metal dichalcogenides: a superior platform for photocatalytic water splitting. \emph{Journal of Physics: Materials} \textbf{2020}, \emph{3}, 022004\relax
\mciteBstWouldAddEndPuncttrue
\mciteSetBstMidEndSepPunct{\mcitedefaultmidpunct}
{\mcitedefaultendpunct}{\mcitedefaultseppunct}\relax
\EndOfBibitem
\bibitem[Y. and et~al.(2021)Y., and et~al.]{65a}
Y.,~L.; et~al.,~F. J. M.~X. Enhanced bulk photovoltaic effect in two-dimensional ferroelectric CuInP$_2$S$_6$. \emph{Nat Commun.} \textbf{2021}, \emph{12}, 5896\relax
\mciteBstWouldAddEndPuncttrue
\mciteSetBstMidEndSepPunct{\mcitedefaultmidpunct}
{\mcitedefaultendpunct}{\mcitedefaultseppunct}\relax
\EndOfBibitem
\bibitem[Chang \latin{et~al.}(2023)Chang, Nanae, Kitamura, Nishimura, Wang, Xiang, Shinokita, Matsuda, Taniguchi, Watanabe, and Nagashio]{61}
Chang,~Y.-R.; Nanae,~R.; Kitamura,~S.; Nishimura,~T.; Wang,~H.; Xiang,~Y.; Shinokita,~K.; Matsuda,~K.; Taniguchi,~T.; Watanabe,~K.; Nagashio,~K. Shift-Current Photovoltaics Based on a Non-Centrosymmetric Phase in In-Plane Ferroelectric SnS. \emph{Advanced Materials} \textbf{2023}, \emph{35}, 2301172\relax
\mciteBstWouldAddEndPuncttrue
\mciteSetBstMidEndSepPunct{\mcitedefaultmidpunct}
{\mcitedefaultendpunct}{\mcitedefaultseppunct}\relax
\EndOfBibitem
\bibitem[Young and Rappe(2012)Young, and Rappe]{77}
Young,~S.~M.; Rappe,~A.~M. First Principles Calculation of the Shift Current Photovoltaic Effect in Ferroelectrics. \emph{Phys. Rev. Lett.} \textbf{2012}, \emph{109}, 116601\relax
\mciteBstWouldAddEndPuncttrue
\mciteSetBstMidEndSepPunct{\mcitedefaultmidpunct}
{\mcitedefaultendpunct}{\mcitedefaultseppunct}\relax
\EndOfBibitem
\bibitem[et~al.(2021)]{77_a}
et~al.,~X. H. W. H. Z.~J. Colossal switchable photocurrents in topological Janus transition metal dichalcogenides. \emph{npj Comput Mater} \textbf{2021}, \emph{7}, 31\relax
\mciteBstWouldAddEndPuncttrue
\mciteSetBstMidEndSepPunct{\mcitedefaultmidpunct}
{\mcitedefaultendpunct}{\mcitedefaultseppunct}\relax
\EndOfBibitem
\bibitem[Pusch \latin{et~al.}(2023)Pusch, R\"omer, Culcer, and Ekins-Daukes]{79}
Pusch,~A.; R\"omer,~U.; Culcer,~D.; Ekins-Daukes,~N.~J. Energy Conversion Efficiency of the Bulk Photovoltaic Effect. \emph{PRX Energy} \textbf{2023}, \emph{2}, 013006\relax
\mciteBstWouldAddEndPuncttrue
\mciteSetBstMidEndSepPunct{\mcitedefaultmidpunct}
{\mcitedefaultendpunct}{\mcitedefaultseppunct}\relax
\EndOfBibitem
\bibitem[Pal \latin{et~al.}(2021)Pal, Muthukrishnan, Sadhukhan, N.~V., Murali, and Murugavel]{74}
Pal,~S.; Muthukrishnan,~S.; Sadhukhan,~B.; N.~V.,~S.; Murali,~D.; Murugavel,~P. {Bulk photovoltaic effect in BaTiO3-based ferroelectric oxides: An experimental and theoretical study}. \emph{Journal of Applied Physics} \textbf{2021}, \emph{129}, 084106\relax
\mciteBstWouldAddEndPuncttrue
\mciteSetBstMidEndSepPunct{\mcitedefaultmidpunct}
{\mcitedefaultendpunct}{\mcitedefaultseppunct}\relax
\EndOfBibitem
\bibitem[Shanyuan \latin{et~al.}(2018)Shanyuan, Debarghya, Kristopher, Yucheng, Yuwei, Elisabeth, Huaixun, B., E., Ralf, R., J., A., Rehan, and Jayakanth]{62}
Shanyuan,~N.; Debarghya,~S.; Kristopher,~W.; Yucheng,~Z.; Yuwei,~L.; Elisabeth,~B.; Huaixun,~H.; B.,~C.~S.; E.,~M.~M.; Ralf,~H.; R.,~J.; J.,~S.~D.; A.,~T.~W.; Rehan,~K.; Jayakanth,~R. Optimal Bandgap in a 2D Ruddlesden–Popper Perovskite Chalcogenide for Single-Junction Solar Cells. \emph{Chemistry of Materials} \textbf{2018}, \emph{30}, 4882--4886\relax
\mciteBstWouldAddEndPuncttrue
\mciteSetBstMidEndSepPunct{\mcitedefaultmidpunct}
{\mcitedefaultendpunct}{\mcitedefaultseppunct}\relax
\EndOfBibitem
\bibitem[Bihari~Swain \latin{et~al.}(2019)Bihari~Swain, Murali, Nanda, and Murugavel]{75}
Bihari~Swain,~A.; Murali,~D.; Nanda,~B.; Murugavel,~P. Large Bulk Photovoltaic Response by Symmetry-Breaking Structural Transformation in Ferroelectric [$\mathrm{Ba}({\mathrm{Zr}}_{0.2}{\mathrm{Ti}}_{0.8}){\mathrm{O}}_{3}$]0.5[(${\mathrm{Ba}}_{0.7}{\mathrm{Ca}}_{0.3}){\mathrm{Ti}\mathrm{O}}_{3}$]0.5. \emph{Phys. Rev. Appl.} \textbf{2019}, \emph{11}, 044007\relax
\mciteBstWouldAddEndPuncttrue
\mciteSetBstMidEndSepPunct{\mcitedefaultmidpunct}
{\mcitedefaultendpunct}{\mcitedefaultseppunct}\relax
\EndOfBibitem
\bibitem[Xiong \latin{et~al.}(2023)Xiong, Chen, Zhang, Cui, Wen, Wen, Wang, Wu, and Sa]{78}
Xiong,~R.; Chen,~X.; Zhang,~Y.; Cui,~Z.; Wen,~J.; Wen,~C.; Wang,~J.; Wu,~B.; Sa,~B. Unraveling the Emerging Photocatalytic, Thermoelectric, and Topological Properties of Intercalated Architecture MZX (M = Ga and In; Z = Si, Ge and Sn; X = S, Se, and Te) Monolayers. \emph{Langmuir} \textbf{2023}, \emph{39}, 0743--7463\relax
\mciteBstWouldAddEndPuncttrue
\mciteSetBstMidEndSepPunct{\mcitedefaultmidpunct}
{\mcitedefaultendpunct}{\mcitedefaultseppunct}\relax
\EndOfBibitem
\bibitem[Deji \latin{et~al.}(2022)Deji, Cedric, Stampfer, and Christoph]{10}
Deji,~L. M. C.~A.; Cedric,~H.; Stampfer; Christoph 2D materials for future heterogeneous electronics. \emph{Nature Communications} \textbf{2022}, \emph{13}, 1392\relax
\mciteBstWouldAddEndPuncttrue
\mciteSetBstMidEndSepPunct{\mcitedefaultmidpunct}
{\mcitedefaultendpunct}{\mcitedefaultseppunct}\relax
\EndOfBibitem
\bibitem[C.W. and et~al.(2022)C.W., and et~al.]{71}
C.W.,~J.; et~al.,~L. W. K.~J. Growth of two-dimensional Janus MoSSe by a single in situ process without initial or follow-up treatments. \emph{NPG Asia Mater} \textbf{2022}, \emph{14}, 15\relax
\mciteBstWouldAddEndPuncttrue
\mciteSetBstMidEndSepPunct{\mcitedefaultmidpunct}
{\mcitedefaultendpunct}{\mcitedefaultseppunct}\relax
\EndOfBibitem
\bibitem[Kresse and Furthm\"uller(1996)Kresse, and Furthm\"uller]{13}
Kresse,~G.; Furthm\"uller,~J. Efficient iterative schemes for ab initio total-energy calculations using a plane-wave basis set. \emph{Physical Review B} \textbf{1996}, \emph{54}, 11169\relax
\mciteBstWouldAddEndPuncttrue
\mciteSetBstMidEndSepPunct{\mcitedefaultmidpunct}
{\mcitedefaultendpunct}{\mcitedefaultseppunct}\relax
\EndOfBibitem
\bibitem[Kresse and Furthmüller(1996)Kresse, and Furthmüller]{38}
Kresse,~G.; Furthmüller,~J. Efficiency of ab-initio total energy calculations for metals and semiconductors using a plane-wave basis set. \emph{Computational Materials Science} \textbf{1996}, \emph{6}, 15--50\relax
\mciteBstWouldAddEndPuncttrue
\mciteSetBstMidEndSepPunct{\mcitedefaultmidpunct}
{\mcitedefaultendpunct}{\mcitedefaultseppunct}\relax
\EndOfBibitem
\bibitem[Perdew \latin{et~al.}(1996)Perdew, Burke, and Ernzerhof]{14}
Perdew,~J.~P.; Burke,~K.; Ernzerhof,~M. Generalized Gradient Approximation Made Simple. \emph{Physical Review Letters} \textbf{1996}, \emph{77}, 3865\relax
\mciteBstWouldAddEndPuncttrue
\mciteSetBstMidEndSepPunct{\mcitedefaultmidpunct}
{\mcitedefaultendpunct}{\mcitedefaultseppunct}\relax
\EndOfBibitem
\bibitem[Blochl(1994)]{15}
Blochl,~P.~E. Projector augmented-wave method. \emph{Physical Review B} \textbf{1994}, \emph{50}, 17953\relax
\mciteBstWouldAddEndPuncttrue
\mciteSetBstMidEndSepPunct{\mcitedefaultmidpunct}
{\mcitedefaultendpunct}{\mcitedefaultseppunct}\relax
\EndOfBibitem
\bibitem[Kresse and Joubert(1999)Kresse, and Joubert]{39}
Kresse,~G.; Joubert,~D. From ultrasoft pseudopotentials to the projector augmented-wave method. \emph{Phys. Rev. B} \textbf{1999}, \emph{59}, 1758--1775\relax
\mciteBstWouldAddEndPuncttrue
\mciteSetBstMidEndSepPunct{\mcitedefaultmidpunct}
{\mcitedefaultendpunct}{\mcitedefaultseppunct}\relax
\EndOfBibitem
\bibitem[Monkhorst and Pack(1976)Monkhorst, and Pack]{16}
Monkhorst,~H.~J.; Pack,~J.~D. Special points for Brillouin-zone integrations. \emph{Physical Review B} \textbf{1976}, \emph{13}, 5188\relax
\mciteBstWouldAddEndPuncttrue
\mciteSetBstMidEndSepPunct{\mcitedefaultmidpunct}
{\mcitedefaultendpunct}{\mcitedefaultseppunct}\relax
\EndOfBibitem
\bibitem[Heyd \latin{et~al.}(2003)Heyd, Scuseria, and Ernzerhof]{81}
Heyd,~J.; Scuseria,~G.~E.; Ernzerhof,~M. {Hybrid functionals based on a screened Coulomb potential}. \emph{The Journal of Chemical Physics} \textbf{2003}, \emph{118}, 8207--8215\relax
\mciteBstWouldAddEndPuncttrue
\mciteSetBstMidEndSepPunct{\mcitedefaultmidpunct}
{\mcitedefaultendpunct}{\mcitedefaultseppunct}\relax
\EndOfBibitem
\bibitem[Gajdo\ifmmode~\check{s}\else \v{s}\fi{} \latin{et~al.}(2006)Gajdo\ifmmode~\check{s}\else \v{s}\fi{}, Hummer, Kresse, Furthm\"uller, and Bechstedt]{76}
Gajdo\ifmmode~\check{s}\else \v{s}\fi{},~M.; Hummer,~K.; Kresse,~G.; Furthm\"uller,~J.; Bechstedt,~F. Linear optical properties in the projector-augmented wave methodology. \emph{Phys. Rev. B} \textbf{2006}, \emph{73}, 045112\relax
\mciteBstWouldAddEndPuncttrue
\mciteSetBstMidEndSepPunct{\mcitedefaultmidpunct}
{\mcitedefaultendpunct}{\mcitedefaultseppunct}\relax
\EndOfBibitem
\bibitem[Wang \latin{et~al.}(2021)Wang, Xu, Liu, Tang, and Geng]{17}
Wang,~V.; Xu,~N.; Liu,~J.-C.; Tang,~G.; Geng,~W.-T. VASPKIT: A user-friendly interface facilitating high-throughput computing and analysis using VASP code. \emph{Computer Physics Communications} \textbf{2021}, \emph{267}, 108033\relax
\mciteBstWouldAddEndPuncttrue
\mciteSetBstMidEndSepPunct{\mcitedefaultmidpunct}
{\mcitedefaultendpunct}{\mcitedefaultseppunct}\relax
\EndOfBibitem
\bibitem[Momma and Izumi(2011)Momma, and Izumi]{40}
Momma,~K.; Izumi,~F. VESTA 3 for three-dimensional visualization of crystal, volumetric and morphology data. \emph{Journal of applied crystallography} \textbf{2011}, \emph{44}, 1272--1276\relax
\mciteBstWouldAddEndPuncttrue
\mciteSetBstMidEndSepPunct{\mcitedefaultmidpunct}
{\mcitedefaultendpunct}{\mcitedefaultseppunct}\relax
\EndOfBibitem
\bibitem[Togo \latin{et~al.}(2023)Togo, Chaput, Tadano, and Tanaka]{18}
Togo,~A.; Chaput,~L.; Tadano,~T.; Tanaka,~I. Implementation strategies in phonopy and phono3py. \emph{J. Phys. Condens. Matter} \textbf{2023}, \emph{35}, 353001\relax
\mciteBstWouldAddEndPuncttrue
\mciteSetBstMidEndSepPunct{\mcitedefaultmidpunct}
{\mcitedefaultendpunct}{\mcitedefaultseppunct}\relax
\EndOfBibitem
\bibitem[Togo(2023)]{19}
Togo,~A. First-principles Phonon Calculations with Phonopy and Phono3py. \emph{J. Phys. Soc. Jpn.} \textbf{2023}, \emph{92}, 012001\relax
\mciteBstWouldAddEndPuncttrue
\mciteSetBstMidEndSepPunct{\mcitedefaultmidpunct}
{\mcitedefaultendpunct}{\mcitedefaultseppunct}\relax
\EndOfBibitem
\bibitem[Mianzeng \latin{et~al.}(2017)Mianzeng, Shuai, Le, Jingbi, Zhongming, Xinfeng, and Jingbo]{80}
Mianzeng,~Z.; Shuai,~Z.; Le,~H.; Jingbi,~Y.; Zhongming,~W.; Xinfeng,~L.; Jingbo,~L. Large-scale 2D PbI$_2$ monolayers: experimental realization and their indirect band-gap related properties. \emph{Nanoscale} \textbf{2017}, \emph{9}, 3736--3741\relax
\mciteBstWouldAddEndPuncttrue
\mciteSetBstMidEndSepPunct{\mcitedefaultmidpunct}
{\mcitedefaultendpunct}{\mcitedefaultseppunct}\relax
\EndOfBibitem
\bibitem[Xiaoli \latin{et~al.}(2019)Xiaoli, Yu, Liping, Mengyuan, Jinyan, and Yucheng]{67}
Xiaoli,~Z.; Yu,~C.; Liping,~S.; Mengyuan,~L.; Jinyan,~D.; Yucheng,~H. Stabilities{,} and electronic and piezoelectric properties of two-dimensional tin dichalcogenide derived Janus monolayers. \emph{J. Mater. Chem. C} \textbf{2019}, \emph{7}, 13203--13210\relax
\mciteBstWouldAddEndPuncttrue
\mciteSetBstMidEndSepPunct{\mcitedefaultmidpunct}
{\mcitedefaultendpunct}{\mcitedefaultseppunct}\relax
\EndOfBibitem
\bibitem[Vallinayagam \latin{et~al.}(2023)Vallinayagam, Sudheer, Aravindh, Murali, Raja, Katta, Posselt, and Zschornak]{42}
Vallinayagam,~M.; Sudheer,~A.~E.; Aravindh,~S.~A.; Murali,~D.; Raja,~N.; Katta,~R.; Posselt,~M.; Zschornak,~M. Novel Metalless Chalcogen-Based Janus Layers: A Density Functional Theory Study. \emph{The Journal of Physical Chemistry C} \textbf{2023}, \emph{127}, 17029--17038\relax
\mciteBstWouldAddEndPuncttrue
\mciteSetBstMidEndSepPunct{\mcitedefaultmidpunct}
{\mcitedefaultendpunct}{\mcitedefaultseppunct}\relax
\EndOfBibitem
\bibitem[Project(2020)]{44}
Project,~T.~M. Materials Data on Pb by Materials Project. \textbf{2020}, \relax
\mciteBstWouldAddEndPunctfalse
\mciteSetBstMidEndSepPunct{\mcitedefaultmidpunct}
{}{\mcitedefaultseppunct}\relax
\EndOfBibitem
\bibitem[Project(2020)]{45}
Project,~T.~M. Materials Data on I by Materials Project. \textbf{2020}, \relax
\mciteBstWouldAddEndPunctfalse
\mciteSetBstMidEndSepPunct{\mcitedefaultmidpunct}
{}{\mcitedefaultseppunct}\relax
\EndOfBibitem
\bibitem[Project(2020)]{46}
Project,~T.~M. Materials Data on Br by Materials Project. \textbf{2020}, \relax
\mciteBstWouldAddEndPunctfalse
\mciteSetBstMidEndSepPunct{\mcitedefaultmidpunct}
{}{\mcitedefaultseppunct}\relax
\EndOfBibitem
\bibitem[Project(2020)]{46b}
Project,~T.~M. Materials Data on Cl2 by Materials Project. \textbf{2020}, \relax
\mciteBstWouldAddEndPunctfalse
\mciteSetBstMidEndSepPunct{\mcitedefaultmidpunct}
{}{\mcitedefaultseppunct}\relax
\EndOfBibitem
\bibitem[Pauling \latin{et~al.}(1970)Pauling, Keaveny, and Robinson]{46a}
Pauling,~L.; Keaveny,~I.; Robinson,~A.~B. The crystal structure of α-fluorine. \emph{Journal of Solid State Chemistry} \textbf{1970}, \emph{2}, 225--227\relax
\mciteBstWouldAddEndPuncttrue
\mciteSetBstMidEndSepPunct{\mcitedefaultmidpunct}
{\mcitedefaultendpunct}{\mcitedefaultseppunct}\relax
\EndOfBibitem
\bibitem[Jain \latin{et~al.}(2013)Jain, Ong, Hautier, Chen, Richards, Dacek, Cholia, Gunter, Skinner, Ceder, and Persson]{43}
Jain,~A.; Ong,~S.~P.; Hautier,~G.; Chen,~W.; Richards,~W.~D.; Dacek,~S.; Cholia,~S.; Gunter,~D.; Skinner,~D.; Ceder,~G.; Persson,~K.~A. {Commentary: The Materials Project: A materials genome approach to accelerating materials innovation}. \emph{APL Materials} \textbf{2013}, \emph{1}, 011002\relax
\mciteBstWouldAddEndPuncttrue
\mciteSetBstMidEndSepPunct{\mcitedefaultmidpunct}
{\mcitedefaultendpunct}{\mcitedefaultseppunct}\relax
\EndOfBibitem
\bibitem[Huang Y.and Pan YH.and~Yang(2020)]{72}
Huang Y.and Pan YH.and~Yang,~R. e.~a. Universal mechanical exfoliation of large-area 2D crystals. \emph{Nat Commun} \textbf{2020}, \emph{11}\relax
\mciteBstWouldAddEndPuncttrue
\mciteSetBstMidEndSepPunct{\mcitedefaultmidpunct}
{\mcitedefaultendpunct}{\mcitedefaultseppunct}\relax
\EndOfBibitem
\bibitem[Mishra \latin{et~al.}(2022)Mishra, Singh, Sonvane, and Ahuja]{50}
Mishra,~P.; Singh,~D.; Sonvane,~Y.; Ahuja,~R. Bifunctional catalytic activity of 2D boron monochalcogenides BX (X = S, Se, Te). \emph{Materials Today Energy} \textbf{2022}, \emph{27}, 101026\relax
\mciteBstWouldAddEndPuncttrue
\mciteSetBstMidEndSepPunct{\mcitedefaultmidpunct}
{\mcitedefaultendpunct}{\mcitedefaultseppunct}\relax
\EndOfBibitem
\bibitem[Tütüncü \latin{et~al.}()Tütüncü, Altuntaş, Srivastava, and Uğur]{25}
Tütüncü,~H.~M.; Altuntaş,~H.; Srivastava,~G.~P.; Uğur,~G. First-principles study of electronic and dynamical properties of AuAl$_2$. \emph{physica status solidi (c)} \emph{1}, 3027--3030\relax
\mciteBstWouldAddEndPuncttrue
\mciteSetBstMidEndSepPunct{\mcitedefaultmidpunct}
{\mcitedefaultendpunct}{\mcitedefaultseppunct}\relax
\EndOfBibitem
\bibitem[Petri\ifmmode~\acute{c}\else \'{c}\fi{} \latin{et~al.}(2021)Petri\ifmmode~\acute{c}\else \'{c}\fi{}, Kremser, Barbone, Qin, Sayyad, Shen, Tongay, Finley, Botello-M\'endez, and M\"uller]{51}
Petri\ifmmode~\acute{c}\else \'{c}\fi{},~M.~M.; Kremser,~M.; Barbone,~M.; Qin,~Y.; Sayyad,~Y.; Shen,~Y.; Tongay,~S.; Finley,~J.~J.; Botello-M\'endez,~A.~R.; M\"uller,~K. Raman spectrum of Janus transition metal dichalcogenide monolayers WSSe and MoSSe. \emph{Phys. Rev. B} \textbf{2021}, \emph{103}, 035414\relax
\mciteBstWouldAddEndPuncttrue
\mciteSetBstMidEndSepPunct{\mcitedefaultmidpunct}
{\mcitedefaultendpunct}{\mcitedefaultseppunct}\relax
\EndOfBibitem
\bibitem[Jiang(2014)]{52}
Jiang,~J.-W. Phonon bandgap engineering of strained monolayer MoS$_2$. \emph{Nanoscale} \textbf{2014}, \emph{6}, 8326--8333\relax
\mciteBstWouldAddEndPuncttrue
\mciteSetBstMidEndSepPunct{\mcitedefaultmidpunct}
{\mcitedefaultendpunct}{\mcitedefaultseppunct}\relax
\EndOfBibitem
\bibitem[Liu \latin{et~al.}(2020)Liu, Wang, Gao, and Li]{53}
Liu,~G.; Wang,~H.; Gao,~Z.; Li,~G.-L. Comparative investigation of the thermal transport properties of Janus SnSSe and SnS2 monolayers. \emph{Phys. Chem. Chem. Phys.} \textbf{2020}, \emph{22}, 16796--16803\relax
\mciteBstWouldAddEndPuncttrue
\mciteSetBstMidEndSepPunct{\mcitedefaultmidpunct}
{\mcitedefaultendpunct}{\mcitedefaultseppunct}\relax
\EndOfBibitem
\bibitem[Matta \latin{et~al.}(2023)Matta, Liao, and Russo]{60}
Matta,~S.~K.; Liao,~T.; Russo,~S.~P. New Janus structure photocatalyst having widely tunable electronic and optical properties with strain engineering. \emph{J. mater. sci. technol.} \textbf{2023}, \emph{155}, 142--147\relax
\mciteBstWouldAddEndPuncttrue
\mciteSetBstMidEndSepPunct{\mcitedefaultmidpunct}
{\mcitedefaultendpunct}{\mcitedefaultseppunct}\relax
\EndOfBibitem
\bibitem[Ravindran \latin{et~al.}(1999)Ravindran, Delin, Johansson, Eriksson, and Wills]{26}
Ravindran,~P.; Delin,~A.; Johansson,~B.; Eriksson,~O.; Wills,~J.~M. Electronic structure, chemical bonding, and optical properties of ferroelectric and antiferroelectric ${\mathrm{NaNO}}_{2}$. \emph{Phys. Rev. B} \textbf{1999}, \emph{59}, 1776--1785\relax
\mciteBstWouldAddEndPuncttrue
\mciteSetBstMidEndSepPunct{\mcitedefaultmidpunct}
{\mcitedefaultendpunct}{\mcitedefaultseppunct}\relax
\EndOfBibitem
\bibitem[Vidya \latin{et~al.}(2002)Vidya, Ravindran, Kjekshus, and Fjellv\aa{}g]{27}
Vidya,~R.; Ravindran,~P.; Kjekshus,~A.; Fjellv\aa{}g,~H. Spin, charge, and orbital ordering in the ferrimagnetic insulator ${\mathrm{YBaMn}}_{2}{\mathrm{O}}_{5}$. \emph{Phys. Rev. B} \textbf{2002}, \emph{65}, 144422\relax
\mciteBstWouldAddEndPuncttrue
\mciteSetBstMidEndSepPunct{\mcitedefaultmidpunct}
{\mcitedefaultendpunct}{\mcitedefaultseppunct}\relax
\EndOfBibitem
\bibitem[Vo \latin{et~al.}(2020)Vo, Vu, Al-Qaisi, Tong, Le, Nguyen, Phuc, Luong, Jappor, Obeid, and Hieu]{59}
Vo,~D.~D.; Vu,~T.~V.; Al-Qaisi,~S.; Tong,~H.~D.; Le,~T.; Nguyen,~C.~V.; Phuc,~H.~V.; Luong,~H.~L.; Jappor,~H.~R.; Obeid,~M.~M.; Hieu,~N.~N. Janus monolayer PtSSe under external electric field and strain: A first principles study on electronic structure and optical properties. \emph{Superlattices and Microstructures} \textbf{2020}, \emph{147}, 106683\relax
\mciteBstWouldAddEndPuncttrue
\mciteSetBstMidEndSepPunct{\mcitedefaultmidpunct}
{\mcitedefaultendpunct}{\mcitedefaultseppunct}\relax
\EndOfBibitem
\bibitem[Chakrapani \latin{et~al.}(2007)Chakrapani, Angus, Anderson, Wolter, Stoner, and Sumanasekera]{82}
Chakrapani,~V.; Angus,~J.; Anderson,~A.; Wolter,~S.; Stoner,~B.; Sumanasekera,~G. Charge Transfer Equilibria Between Diamond and an Aqueous Oxygen Electrochemical Redox Couple. \emph{Science (New York, N.Y.)} \textbf{2007}, \emph{318}, 1424--30\relax
\mciteBstWouldAddEndPuncttrue
\mciteSetBstMidEndSepPunct{\mcitedefaultmidpunct}
{\mcitedefaultendpunct}{\mcitedefaultseppunct}\relax
\EndOfBibitem
\bibitem[Sun and Schwingenschlögl(2020)Sun, and Schwingenschlögl]{48}
Sun,~M.; Schwingenschlögl,~U. B2P6 : A Two-Dimensional Anisotropic Janus Material with Potential in Photocatalytic Water Splitting and Metal-Ion Batteries. \emph{Chemistry of Materials} \textbf{2020}, \relax
\mciteBstWouldAddEndPunctfalse
\mciteSetBstMidEndSepPunct{\mcitedefaultmidpunct}
{}{\mcitedefaultseppunct}\relax
\EndOfBibitem
\bibitem[Zhuang and Hennig(2013)Zhuang, and Hennig]{24}
Zhuang,~H.~L.; Hennig,~R.~G. Theoretical perspective of photocatalytic properties of single-layer {S}n{S}$_{2}$. \emph{Phys. Rev. B} \textbf{2013}, \emph{88}, 115314\relax
\mciteBstWouldAddEndPuncttrue
\mciteSetBstMidEndSepPunct{\mcitedefaultmidpunct}
{\mcitedefaultendpunct}{\mcitedefaultseppunct}\relax
\EndOfBibitem
\bibitem[Toroker \latin{et~al.}(2011)Toroker, Kanan, Alidoust, Isseroff, Liao, and Carter]{41}
Toroker,~M.~C.; Kanan,~D.~K.; Alidoust,~N.; Isseroff,~L.~Y.; Liao,~P.; Carter,~E.~A. First principles scheme to evaluate band edge positions in potential transition metal oxide photocatalysts and photoelectrodes. \emph{Phys. Chem. Chem. Phys.} \textbf{2011}, \emph{13}, 16644--16654\relax
\mciteBstWouldAddEndPuncttrue
\mciteSetBstMidEndSepPunct{\mcitedefaultmidpunct}
{\mcitedefaultendpunct}{\mcitedefaultseppunct}\relax
\EndOfBibitem
\bibitem[Gao \latin{et~al.}(2021)Gao, Shen, Liu, Lv, Zhou, Zhou, Feng, and Shen]{73}
Gao,~X.; Shen,~Y.; Liu,~J.; Lv,~L.; Zhou,~M.; Zhou,~Z.; Feng,~Y.~P.; Shen,~L. Boosting the photon absorption{,} exciton dissociation{,} and photocatalytic hydrogen- and oxygen-evolution reactions by built-in electric fields in Janus platinum dichalcogenides. \emph{J. Mater. Chem. C} \textbf{2021}, \emph{9}, 15026--15033\relax
\mciteBstWouldAddEndPuncttrue
\mciteSetBstMidEndSepPunct{\mcitedefaultmidpunct}
{\mcitedefaultendpunct}{\mcitedefaultseppunct}\relax
\EndOfBibitem
\end{mcitethebibliography}

\end{document}


\begin{figure}
    \centering
    \includegraphics[height =15 cm ]{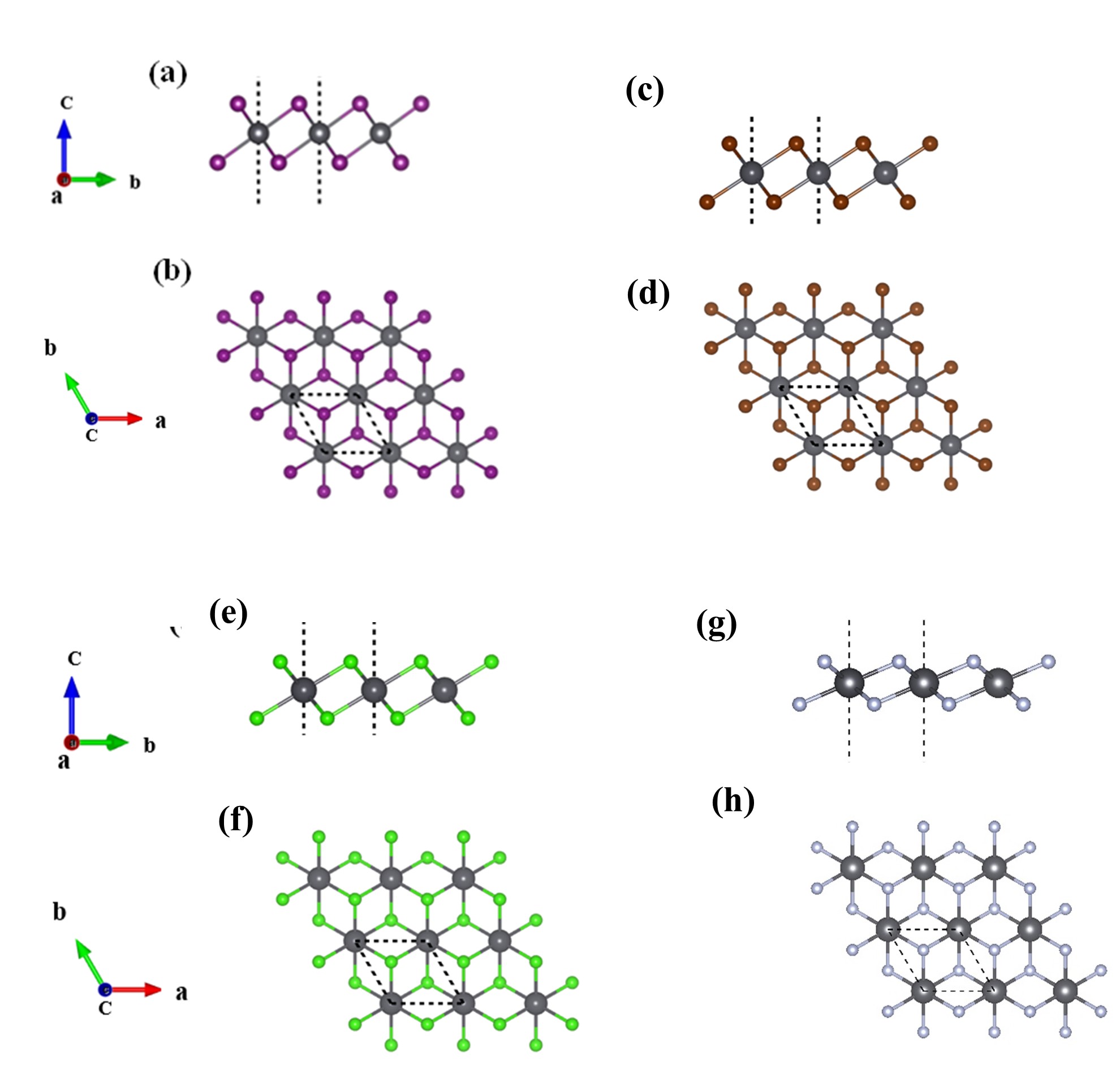}
    \caption{The side and top view of geometrically relaxed PbI$_2$ ((a) and (b)), PbBr$_2$ ((c) and (d)), PbCl$_2$ ((e) and (f)), and PbF$_2$ ((g) and (h)) monolayers. The colour black, violet, brown, green and grey represent Pb, I, Br, Cl and F respectively.}
\end{figure}
\begin{figure}
    \centering
    \includegraphics{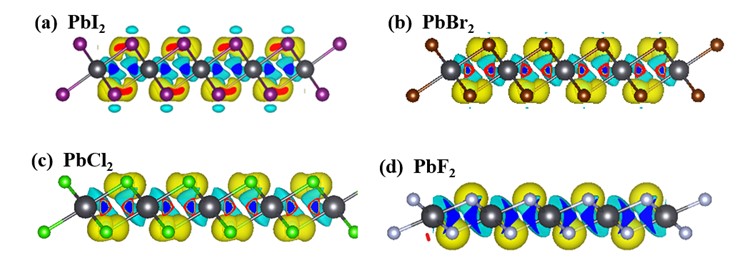}
    \caption{The differential charge density of PbI$_2$ (a), PbBr$_2$ (b), PbCl$_2$ (c) and PbF$_2$ (d)  monolayers were presented at the isosurface level of 0.002 e/\AA$^3$. The blue and yellow region represent charge depleted and charge accumulated region.}
\end{figure}
\begin{figure}[H]
    \centering
    \includegraphics{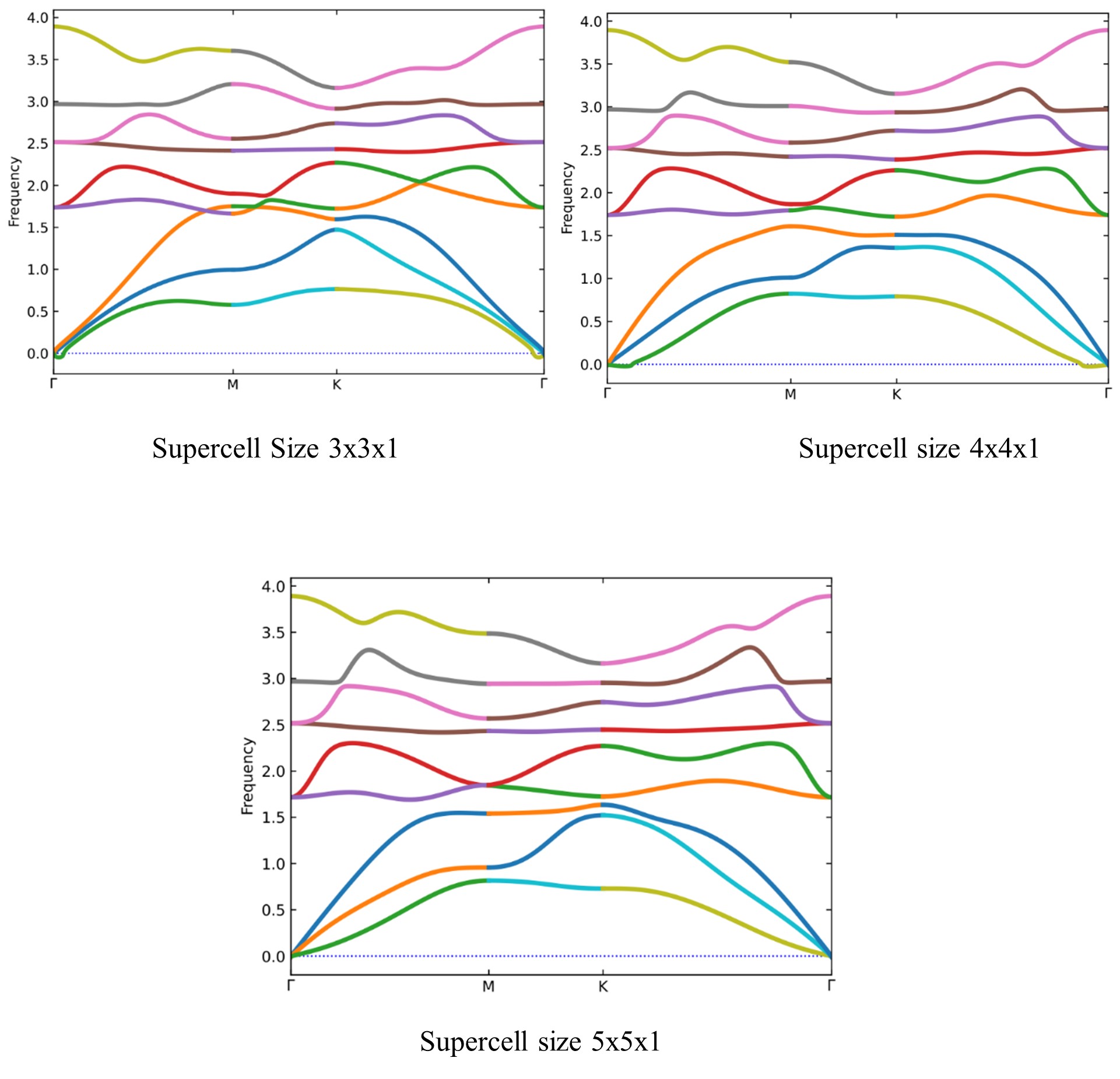}
    \caption{Phonon spectra of PbIBr Janus structure with different supercell size.}
\end{figure}
\begin{figure}
    \centering
    \includegraphics{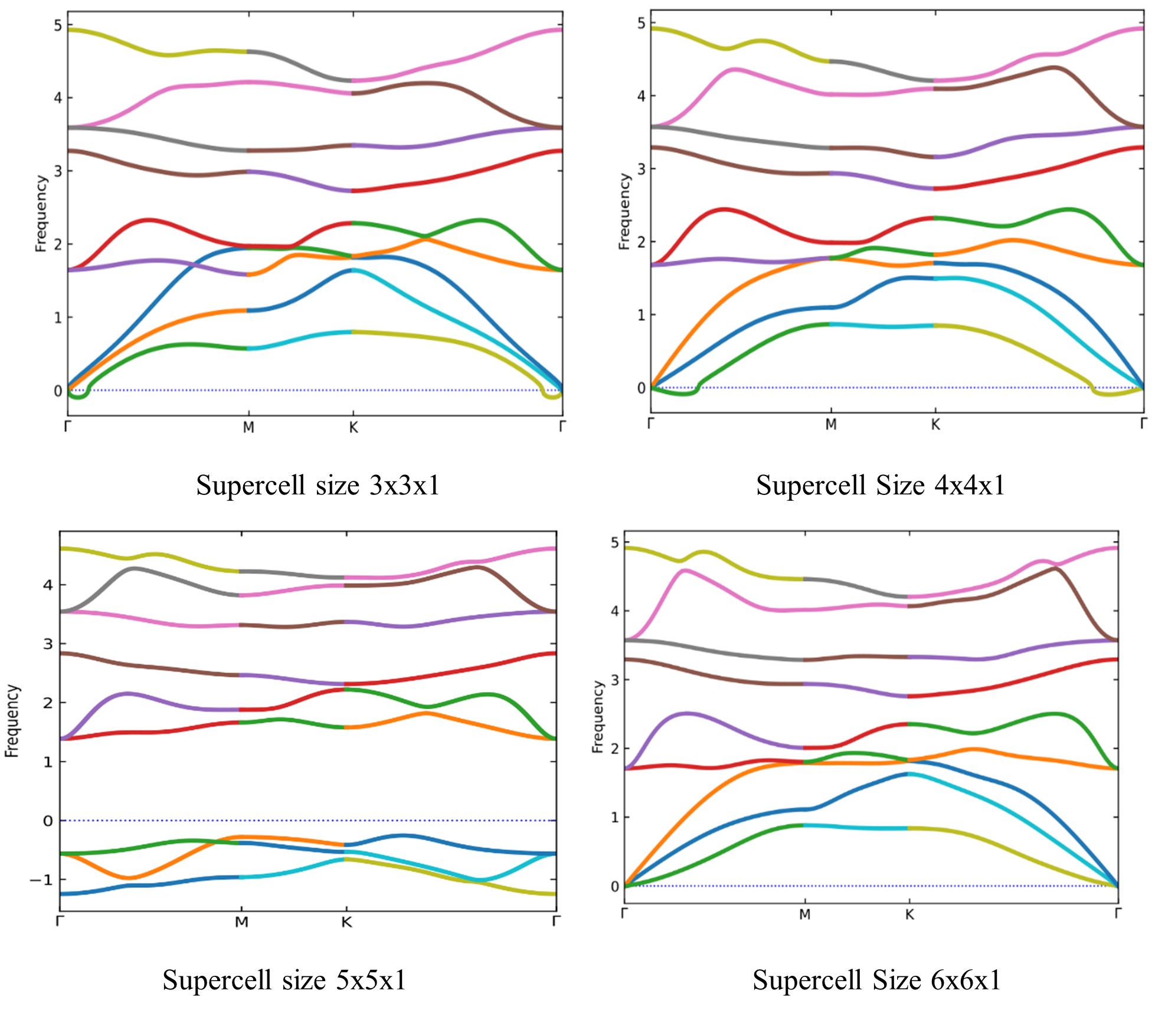}
    \caption{Phonon spectra of PbICl Janus structure with different supercell size.}
\end{figure}
\begin{figure}
    \centering
    \includegraphics{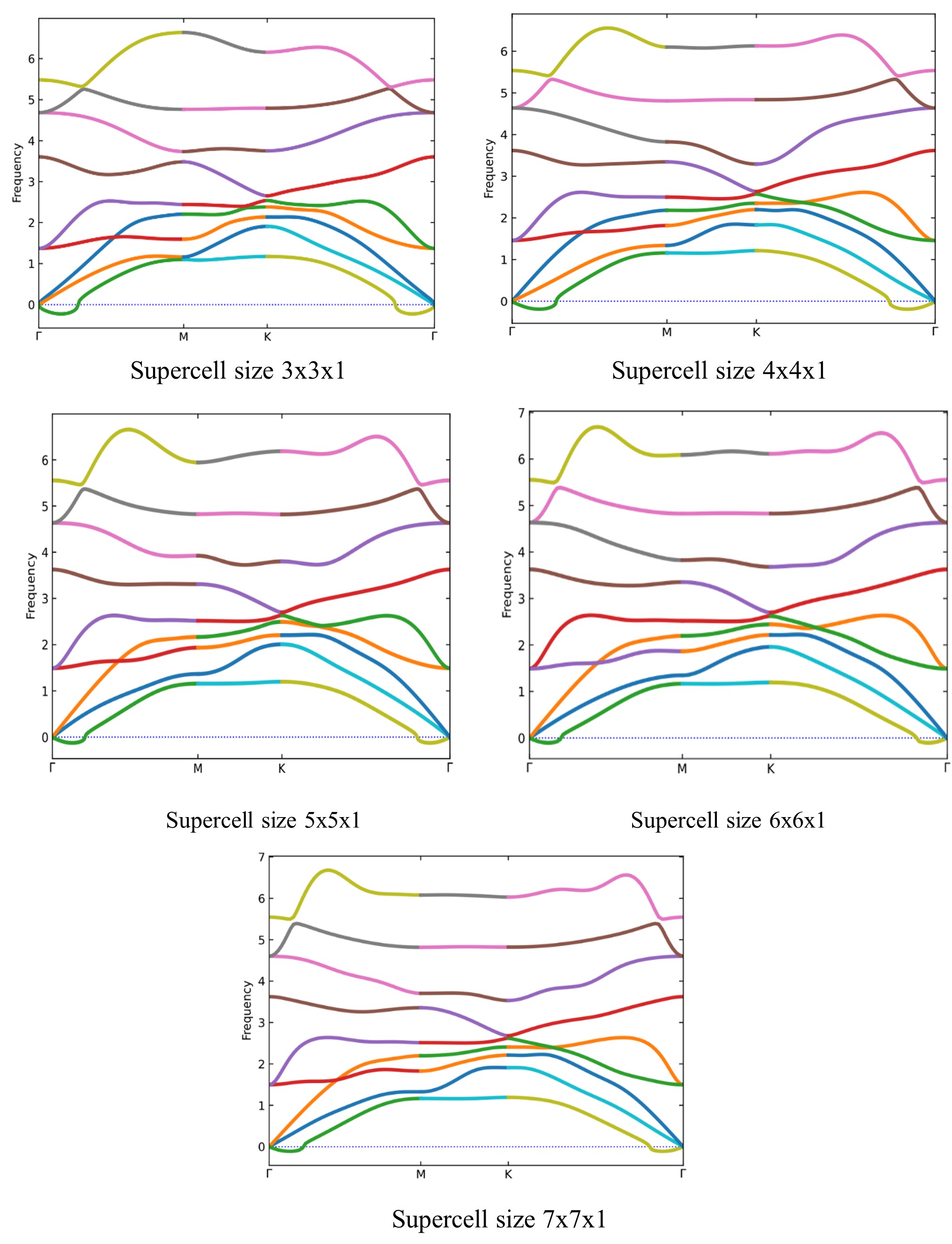}
    \caption{Phonon spectra of PbIF Janus structure with different supercell size.}
\end{figure}
\begin{figure}
    \centering
    \includegraphics{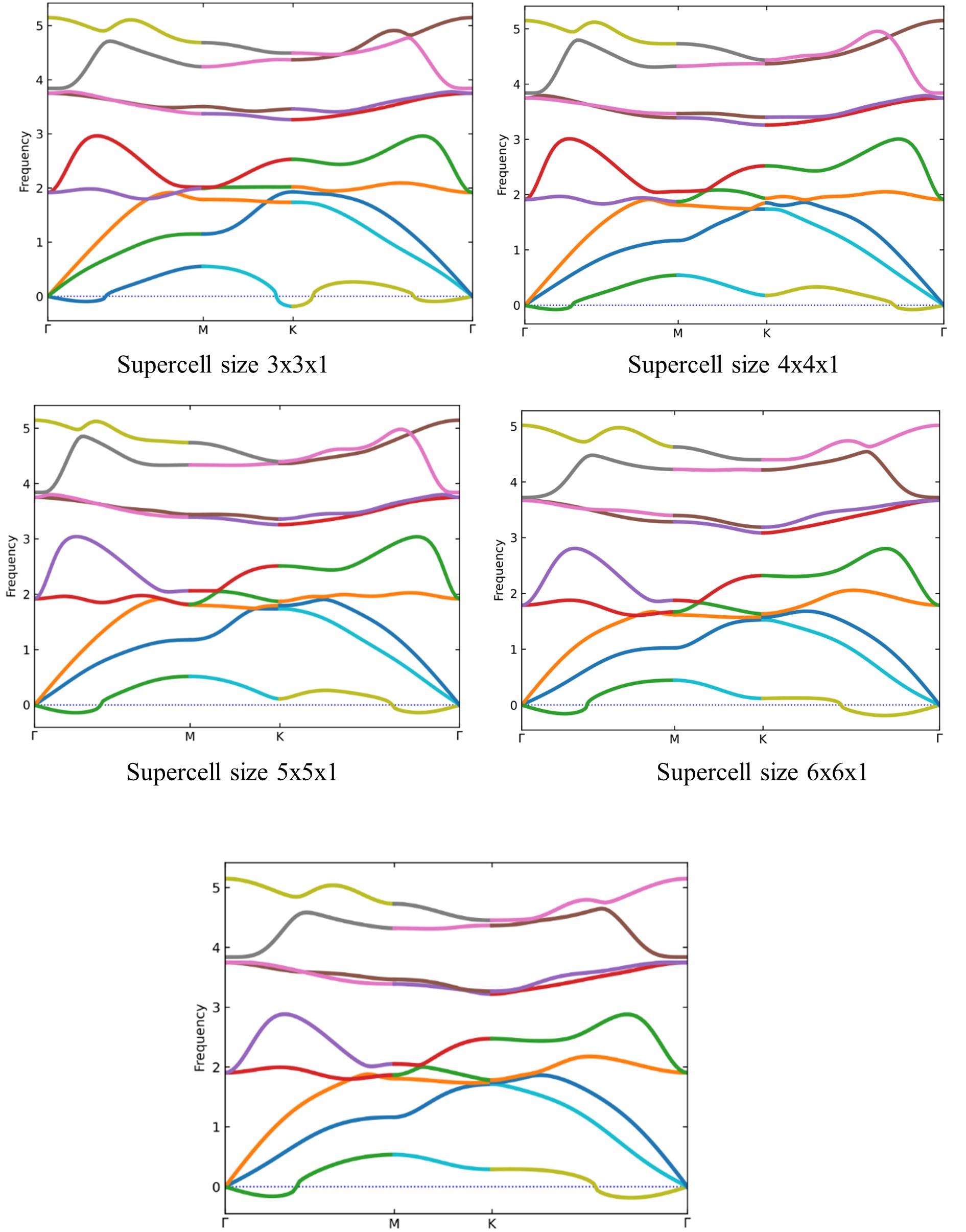}
    \caption{Phonon spectra of PbBrCl Janus structure with different supercell size.}
\end{figure}
\begin{figure}
    \centering
    \includegraphics{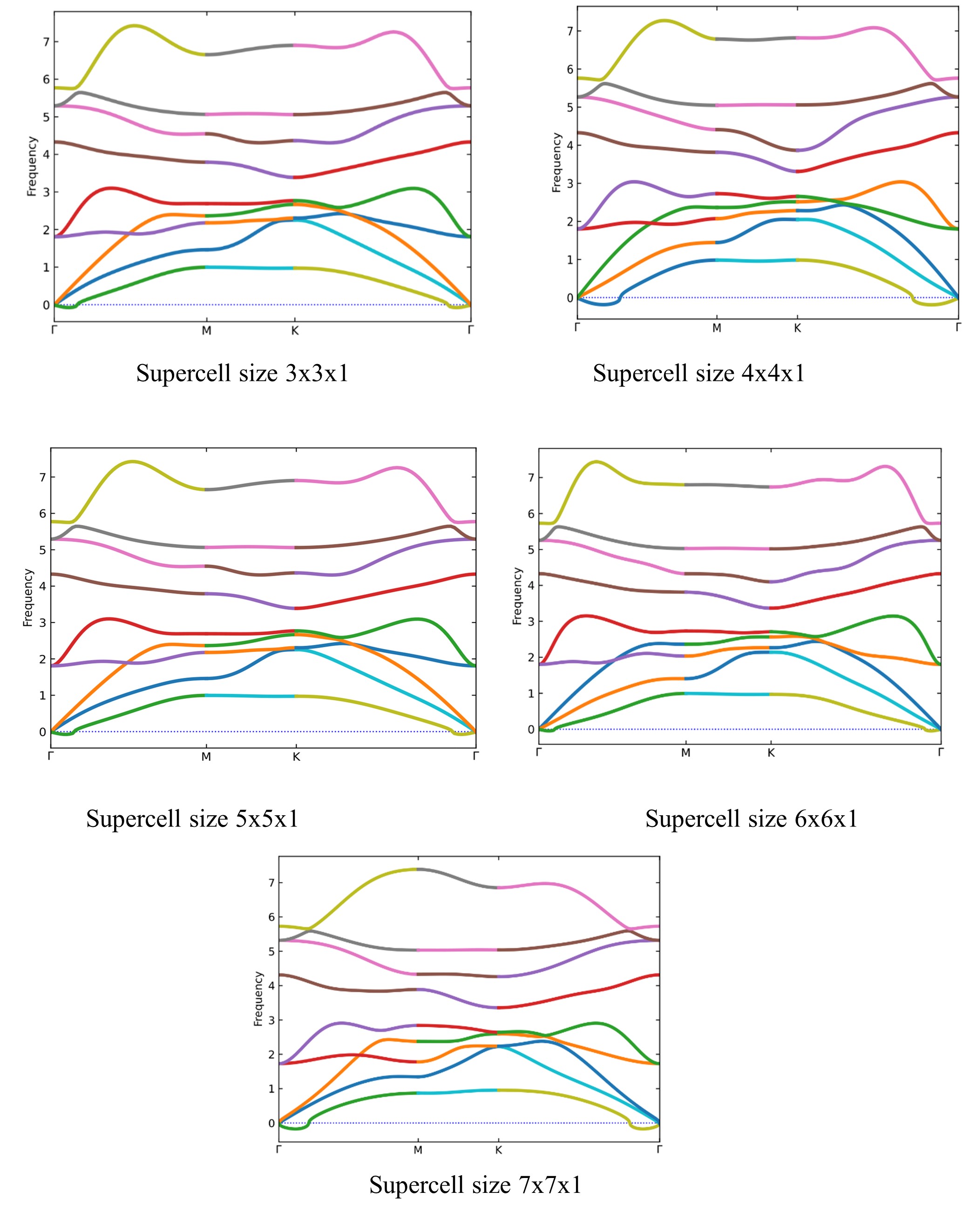}
    \caption{Phonon spectra of PbBrF Janus structure with different supercell size.}
\end{figure}
\begin{figure}
    \centering
    \includegraphics{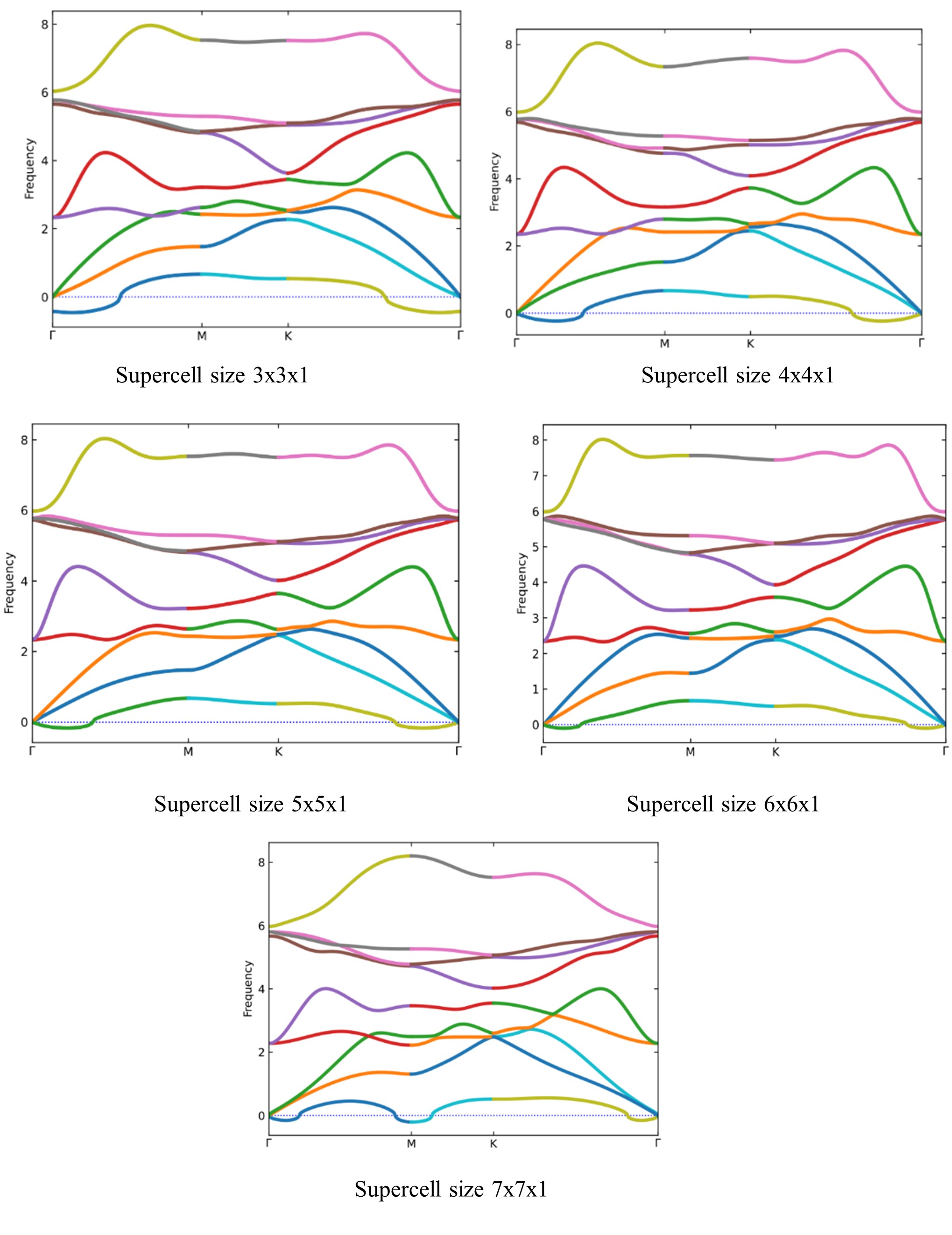}
    \caption{Phonon spectra of PbClF Janus structure with different supercell size.}
\end{figure}
\begin{figure}
    \centering
    \includegraphics{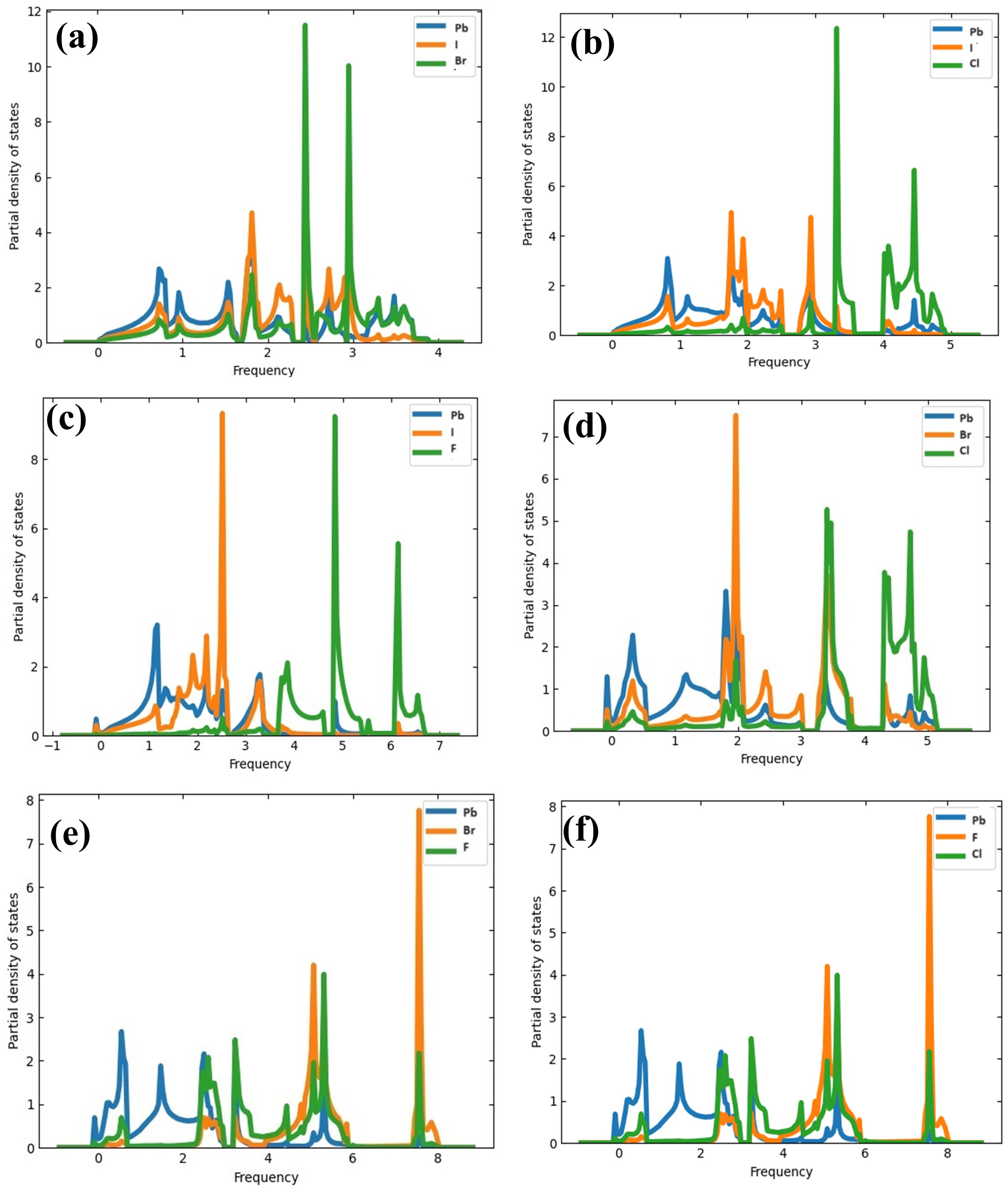}
    \caption{The atom resolved phonon DOS of PbIBr (5x5x1) (a), PbICl (6x6x1) (b), PbIF (7x7x1) (c), PbBrCl (7x7x1) (d), PbBrF (7x7x1) (e), and PbClF (7x7x1) (f), structures. The supercell size is given in bracket}
\end{figure}
\begin{figure}
    \centering
    \includegraphics[height =15 cm]{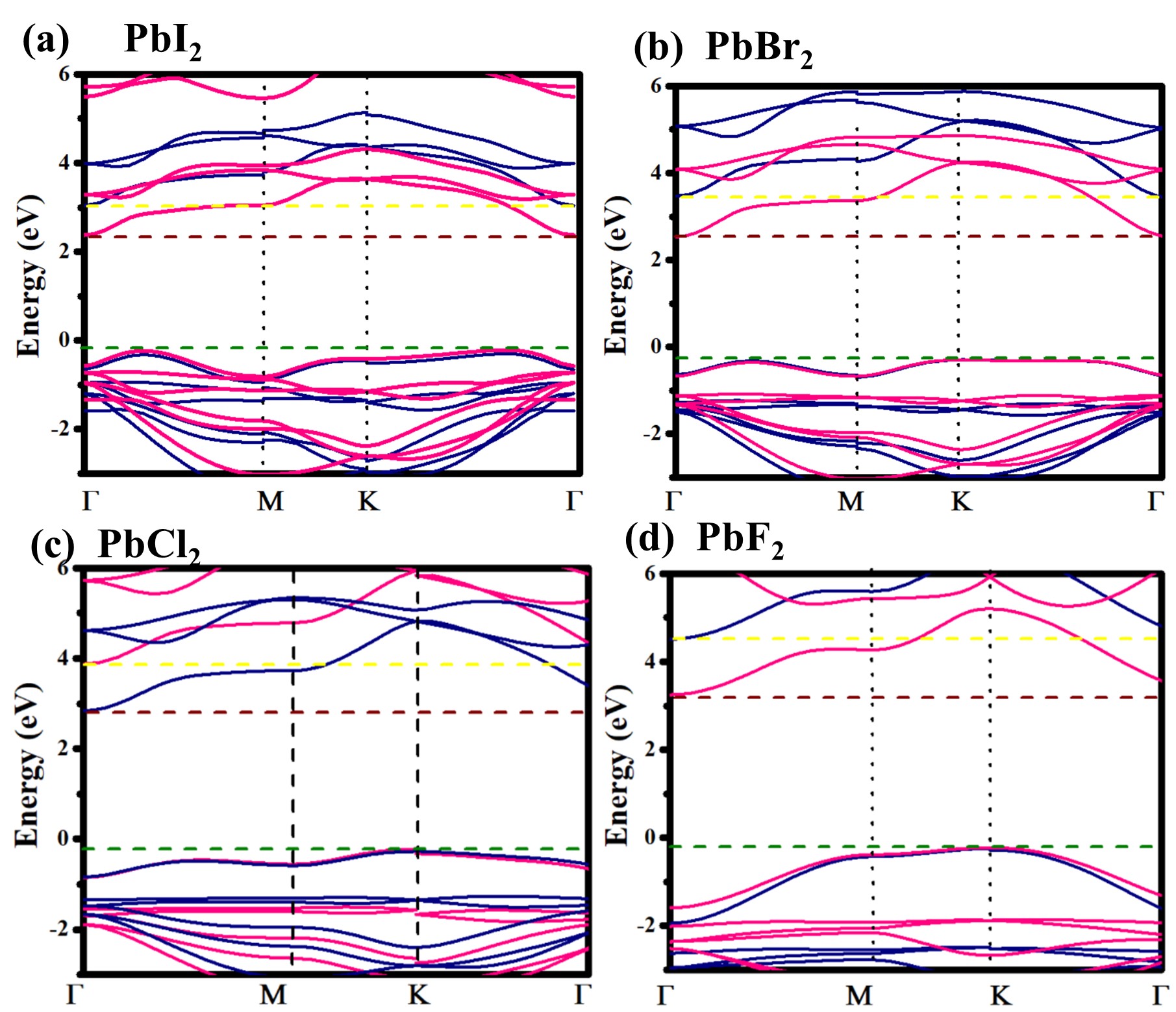}
    \caption{Basic electronic band structure calculated for PbI$_2$ (a), PbBr$_2$ (b), PbCl$_2$ (c) and PbF$_2$ (d) structures using GGA (blue line) and HSE (pink line) level theory. The green dotted line represent valence band maximum (VBM) corresponds to both GGA and HSE. The brown and yellow dotted line represent conduction band minimum (CBM) calculated using GGA and HSE functional.}
\end{figure}
\begin{figure}
    \centering
    \includegraphics[height = 15 cm]{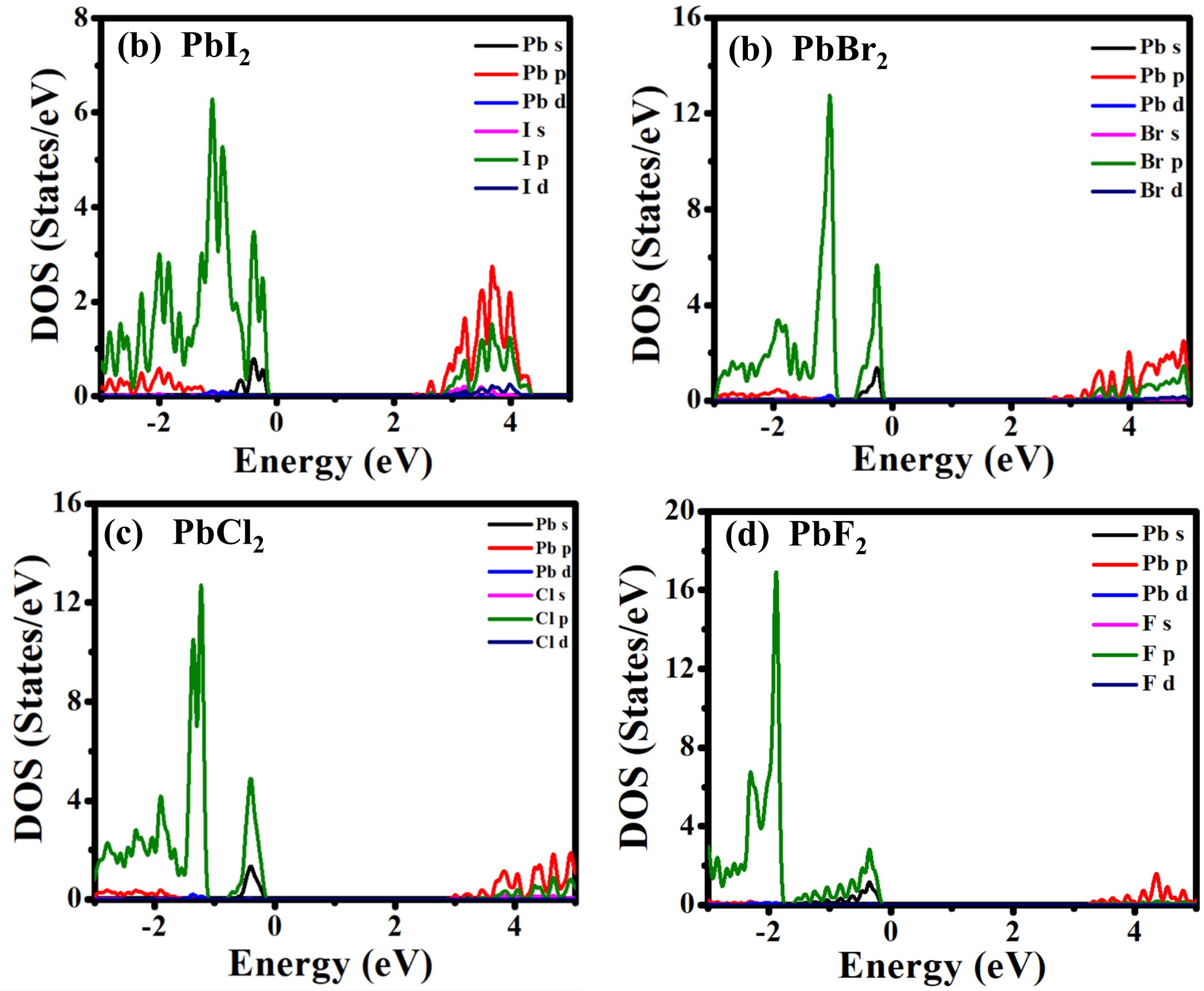}
    \caption{Atom resolved density of states of PbI$_2$ (a), PbBr$_2$ (b), PbCl$_2$ (c) and PbF$_2$ (d) monolayers. CBM is mainly constructed out of p orbital of Pb atom in all the parent structures.}
\end{figure}
\begin{figure}
    \centering
    \includegraphics{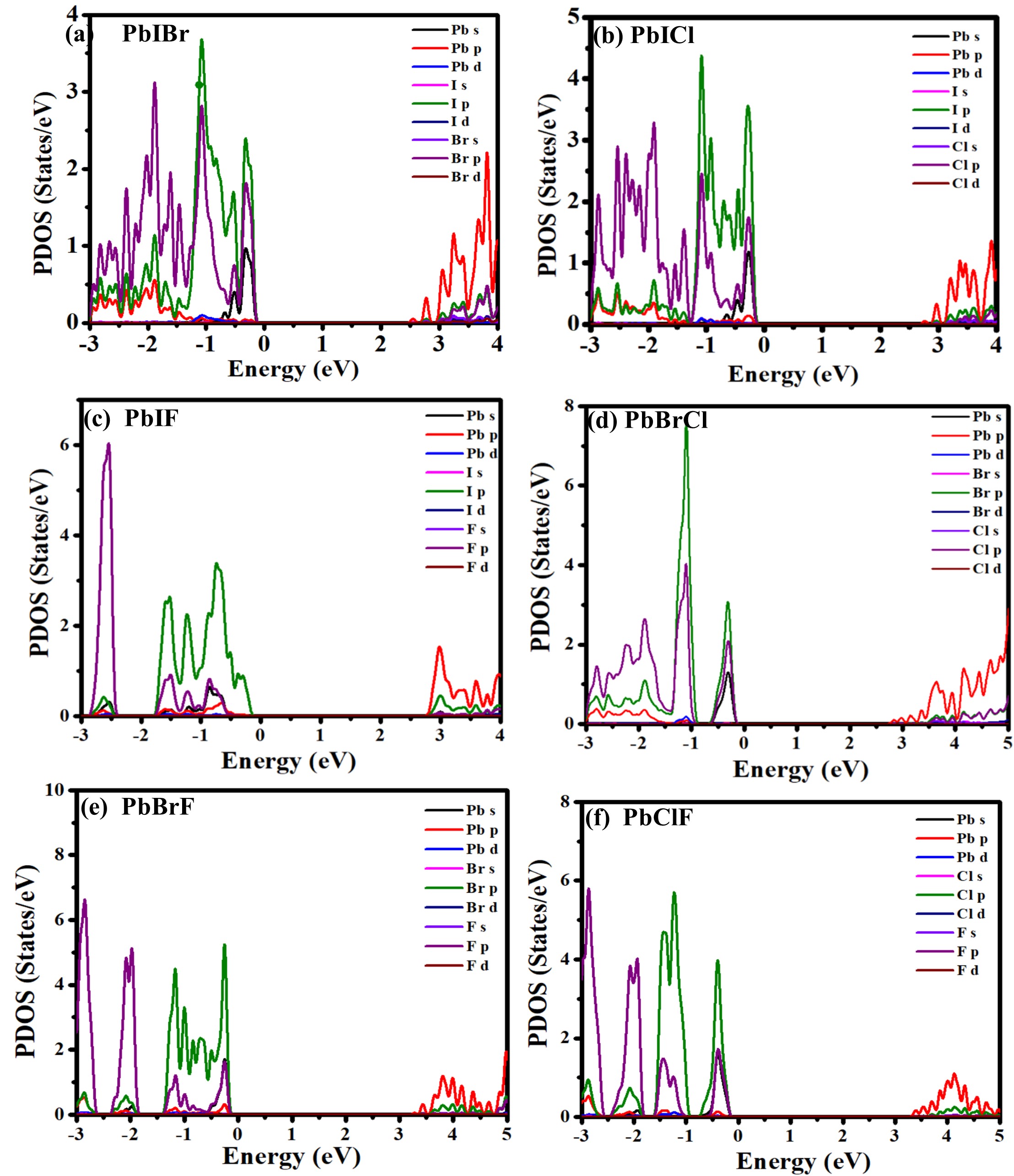}
    \caption{Atom resolved density of states of PbIBr (a), PbICl (b), PbIF (c), PbBrCl (d), PbBrF (e) and PbClF (f) Janus structures.}
\end{figure}
\begin{figure}
    \centering
    \includegraphics{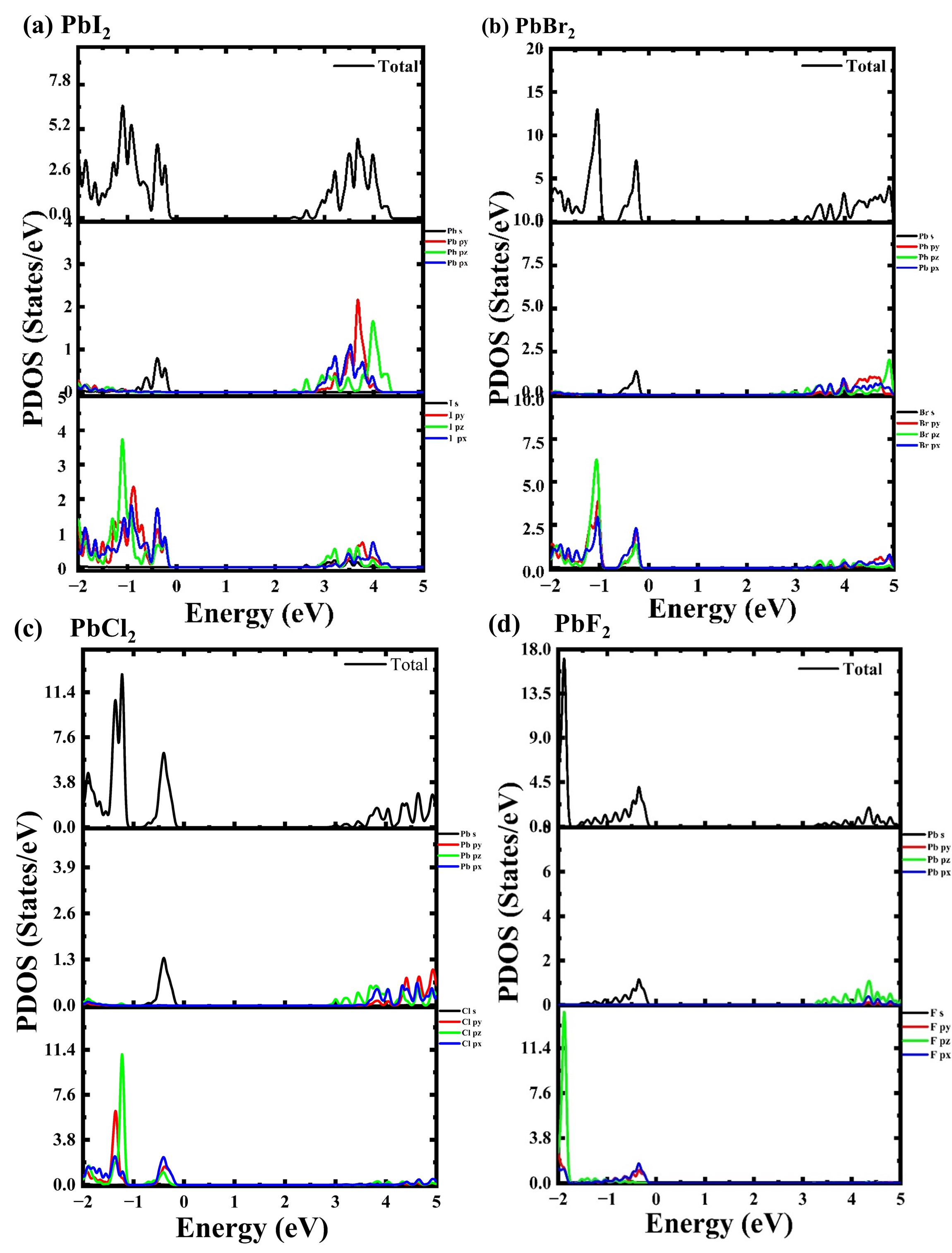}
    \caption{Orbital resolved density of states of PbI$_2$ (a), PbBr$_2$ (b), PbCl$_2$ (c) and PbF$_2$ (d) monolayer structures.}
\end{figure}
\begin{figure}
    \centering
    \includegraphics{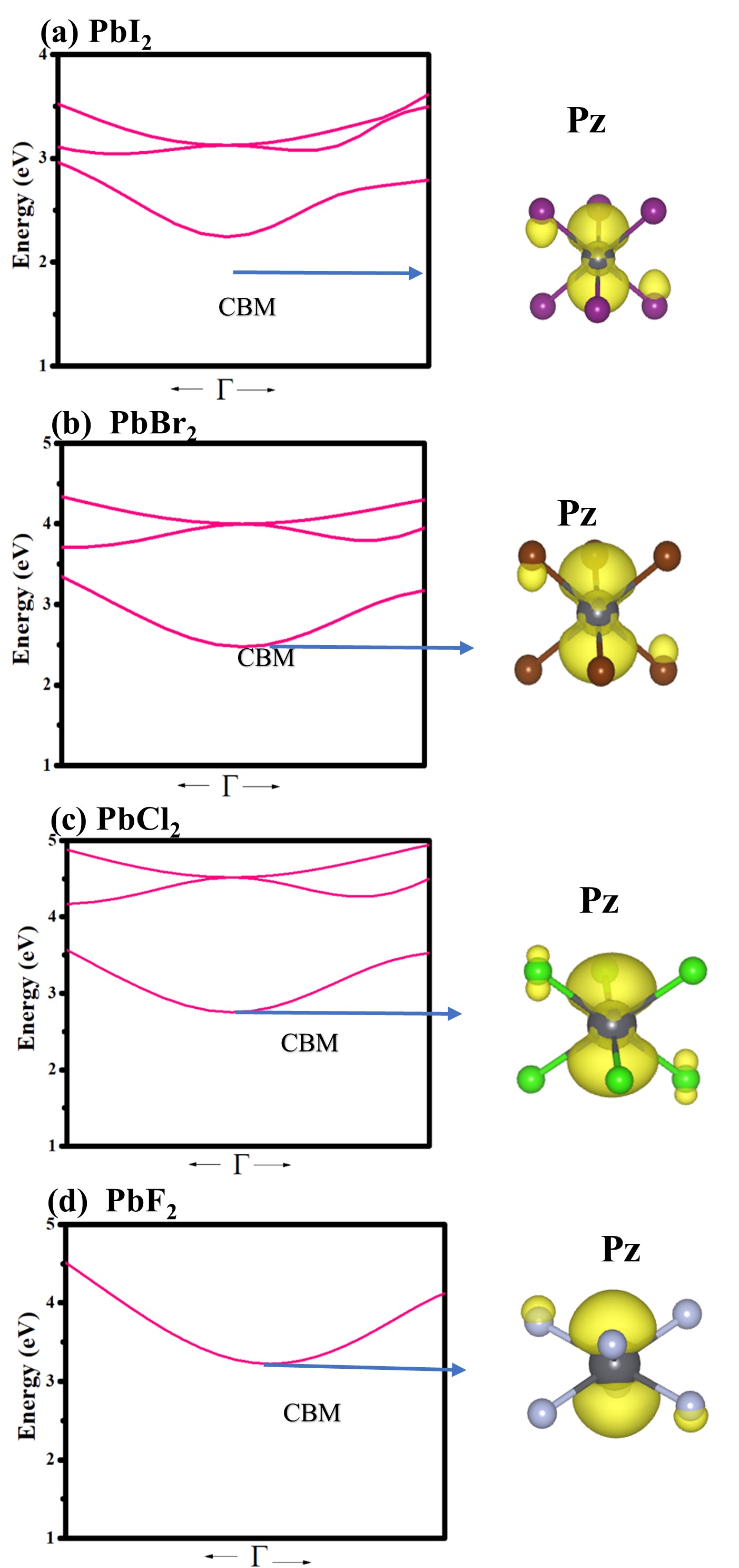}
    \caption{Conduction band minimum and the partial charge density calculated at the CBM of PbI$_2$ (a), PbBr$_2$ (b), PbCl$_2$ (c) and PbF$_2$ (d) monolayers.}
\end{figure}
\begin{figure}
    \centering
    \includegraphics{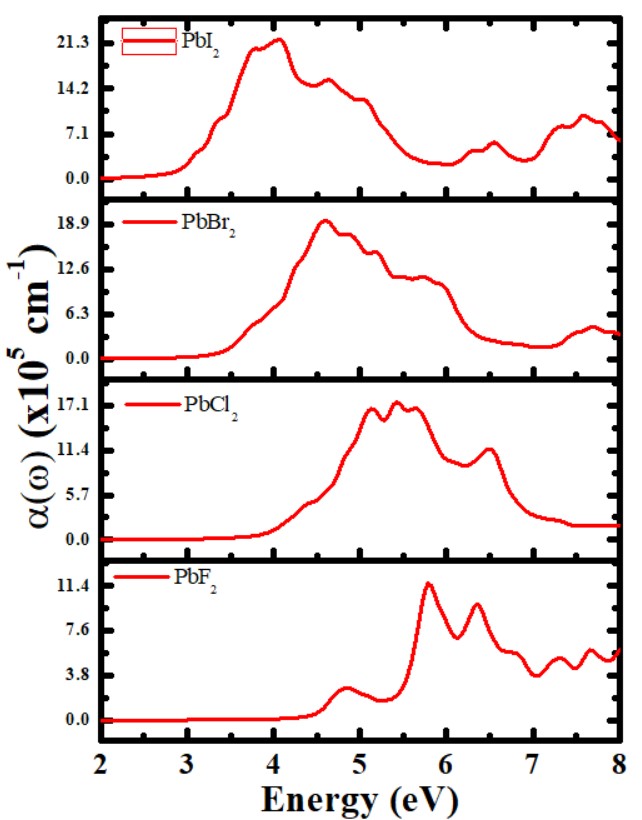}
    \caption{Optical absorption spectra simulated for PbI$_2$, PbBr$_2$, PbCl$_2$ and PbF$_2$ monolayers.}
\end{figure}
\begin{figure}
    \centering
    \includegraphics[height = 16 cm]{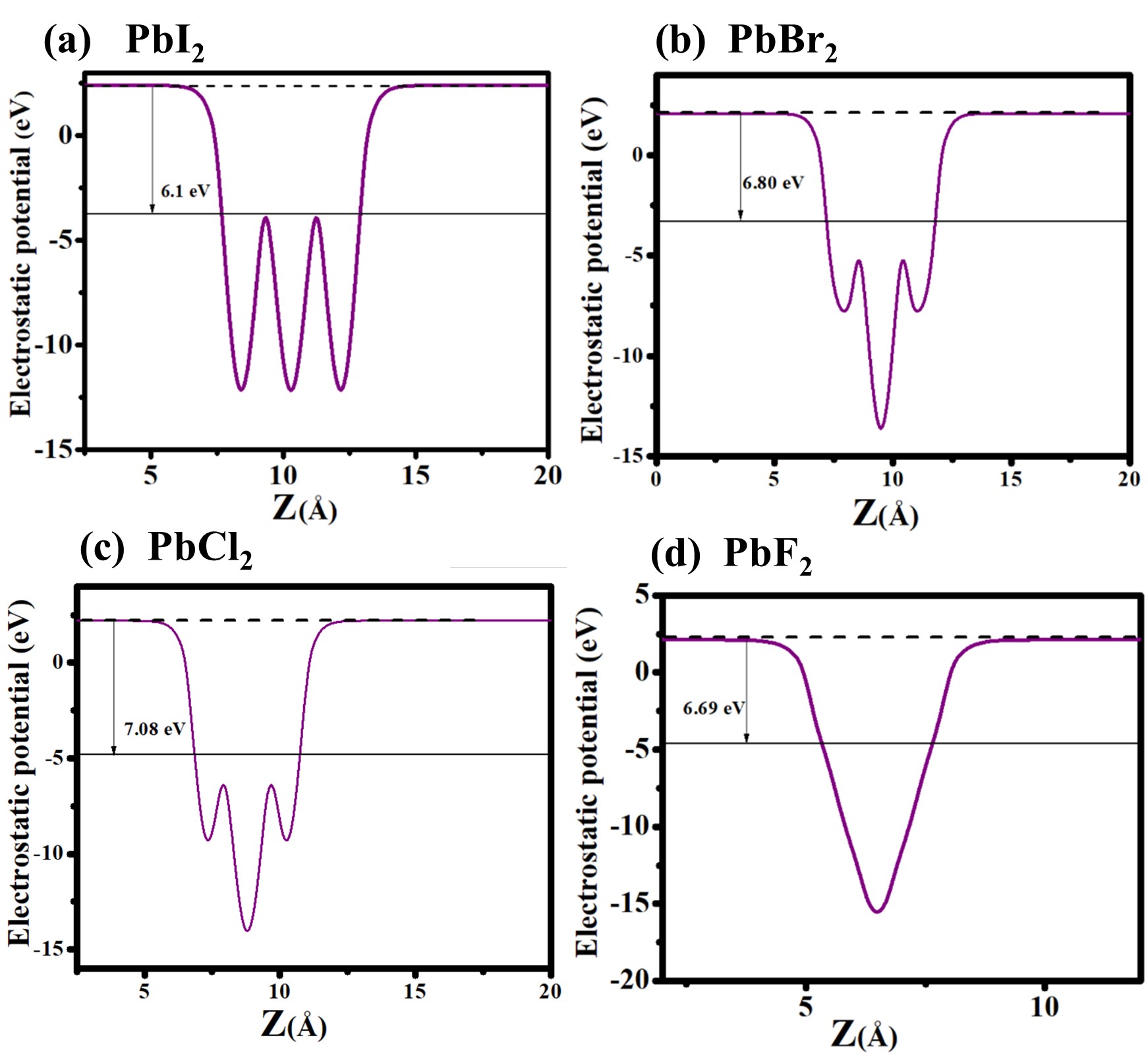}
    \caption{Electrostatic potential diagram of PbI$_2$ (a), PbBr$_2$ (b), PbCl$_2$ (c) and PbF$_2$ (d) monolayers.  The dotted line and the black line represents vacuum level and fermi energy level. }
\end{figure}
\begin{figure}
    \centering
    \includegraphics{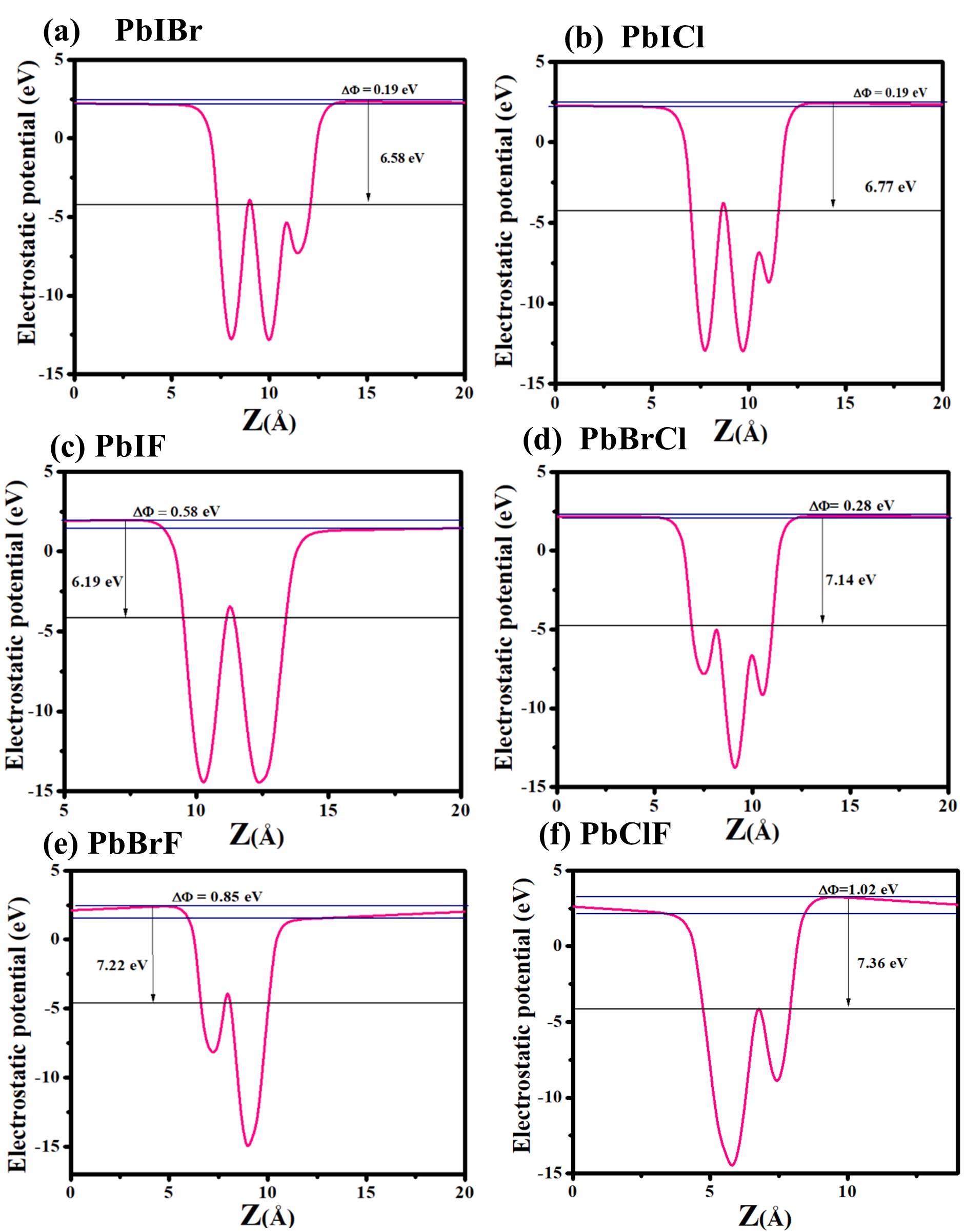}
    \caption{Electrostatic potential diagram of PbIBr (a), PbICl (b), PbIF (c), PbBrCl (d), PbBrF (e), and PbClF (f) monolayers. Here, the black and blue line represent fermi energy level and vacuum level with respect to top and bottom later. This indicate electrostatic potential difference between top and bottom layer.}
\end{figure}
\begin{figure}
    \centering
    \includegraphics{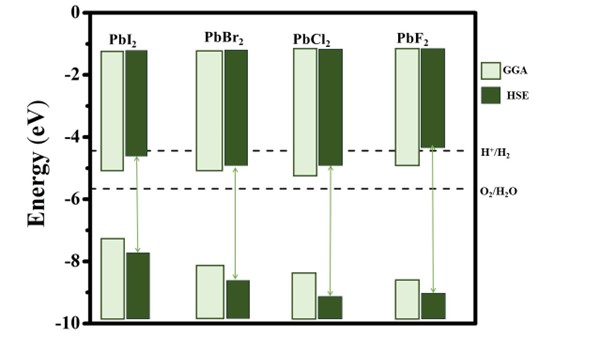}
    \caption{The valence band and the conduction band alignment of PbI$_2$, PbBr$_2$, PbCl$_2$ and PbF$_2$ monolayers. The white and green denote the band alignment correspond to GGA and HSE functional. The dotted line indicates water reduction and oxidation potential at pH equal to zero.}
\end{figure}
\begin{table*}
   \caption{Structural information of PbX$_2$ (X = F, Cl, Br, I) Janus structures.}
   
\renewcommand{\arraystretch}{1}
    \begin{tabular}{ |p{2cm}p{1.9cm}p{4.5cm} p{2cm}p{2cm}|}   
    \hline
\normalsize{Structure} &\normalsize{Lattice parameter (\AA) (a=b) }&\normalsize{Bond length (\AA)}& \normalsize{Bond angle (Degree) (@centerlayer)} & \normalsize{Layer thickness (\AA) }\\[0.5ex]
 \hline\hline
 \hspace{0.5cm}
 \vspace{0.25cm}
\small PbI$_2$ &\small4.55&\small3.25&	\small90.94& \small3.82\\[1ex]
\hspace{0.5cm}\vspace{0.25cm}
\small PbBr$_2$ &\small	4.41&\small 3.06 & \small 87.80 &\small 3.40\\[1ex]
\hspace{0.5cm}\vspace{0.25cm}
\small PbCl$_2$ &\small 4.34 & \small2.93& \small84.39& \small3.04\\[1ex]
\hspace{0.5cm}\vspace{0.25cm}
\small PbF$_2$ & \small3.99 & \small2.50  & \small74.53 &\small 1.98 \\[1ex]
\hline 
\end{tabular}
\end{table*}
\begin{table*}
 \caption{Formation energy and Bader charge analysis of PbXY (X, Y = F, Cl, Br, I) Janus structures.}
\centering
   \begin{tabular}{ ||c c c c c c c  c || }
 \hline

 \hline
Structure & Formation Energy&  Pb (2+)& I (1-)& Br (1-)& Cl (1-)& F (1-)   &\\[1ex] 
 \hline\hline
PbI$_2$ & -2.34& 0.93& -0.46&&&&\\[1ex]
PbBr$_2$ &-2.99& 1.13&& -0.56&&&\\[1ex]
PbCl$_2$  &-3.51&1.28&&&-0.63&&\\[1ex]
PbF$_2$ &-6.96& 1.53& & &&-.78&\\[1ex] 
 \hline
 \end{tabular}  
\end{table*}
\begin{table*}
 \caption{Chemical potential of Pb, I, Br, Cl and F elements are given. The ground state structure of each element is selected from the material project database, having energy above the hull.}
\centering
 \begin{tabular}{ ||c c c || }
 \hline

 \hline
Species& Elemental Structure&Chemical potential (eV)  \\[1ex] 
 \hline\hline
Pb & Cubic(Fm\(\Bar{3}\)m)& -3.56\\[1ex]
I &Orthorhombic (Immm)&-1.57\\[1ex]
Br &Orthorhombic(Immm)&-1.71\\[1ex]
Cl &Orthorhombic(Cmce)&-1.93\\[1ex]
F& Monoclinic(C12/c1)&-1.81\\[1ex]
 \hline
 \end{tabular}  
\end{table*}

\begin{table}
    \begin{tabular}{||p{2cm}p{2cm}p{1.7cm}p{2cm}p{2cm}p{2cm}||} 
  \hline
  Structure& Supercell & Lattice parameter& Number of atoms& Kpoints& Error Percentage\\[1ex]
  \hline
 PbIBr& 3x3x1 & 13.47&27&5x5x1&1 \\ [1ex]
 &4x4x1&17.96&48&5x5x1&0.5 \\[1ex]
&5x5x1&22.45&75&3x3x1&0\\[1ex]
  & 6x6x1&&&& \\ 
 \hline
  PbICl& 3x3x1& 13.36 &27& 5x5x1& 2 \\ [1ex]
 &4x4x1&17.81&48&5x5x1&2.5\\[1ex]
 &5x5x1&22.27&75&3x3x1&10\\[1ex]
 &6x6x1&26.72&108&3x3x1&0\\[1ex]
 \hline
 PbIF&3x3x1&12.73&27&5x5x1&3\\[1ex]
 &4x4x1&16.96&48&5x5x1&3\\[1ex]
 &5x5x1&21.20&75&3x3x1&2\\[1ex]
 &6x6x1&25.44&108&3x3x1&2\\[1ex]
 &7x7x1&29.68&147&3x3x1&3\\[1ex]
 \hline
 PbBrCl&3x3x1&13.13&27&5x5x1&3\\[1ex]
 &4x4x1&17.51&48&5x5x1&3\\[1ex]
 &5x5x1&21.89&75&3x3x1&2\\[1ex]
 &6x6x1&26.27&108&3x3x1&3\\[1ex]
 &7x7x1&30.65&147&3x3x1&3\\[1ex]
 \hline
 PbBrF&3x3x1&12.46&27&5x5x1&2\\[1ex]
 &4x4x1&16.62&48&5x5x1&2\\[1ex]
 &5x5x1&20.78&75&3x3x1&1\\[1ex]
 &6x6x1&24.93&108&3x3x1&0.5\\[1ex]
 &7x7x1&29.09&147&3x3x1&0.25\\[1ex]
 \hline
 PbClF&3x3x1&12.36&27&5x5x1&4\\[1ex]
 &4x4x1&16.49&48&5x5x1&6\\[1ex]
 &5x5x1&20.61&75&3x3x1&4\\[1ex]
 &6x6x1&24.73&108&3x3x1&3\\[1ex]
 &7x7x1&28.86&147&3x3x1&1\\[1ex]
 \hline
\end{tabular}
    \caption{Dynamical stability analysis of PbXY Janus structures with different supercell. Error percentage of imaginary modes in the phonon spectra of Janus structures were also presented.}
\end{table}
 \begin{table*}
    \caption{The basic electronic properties derived from band structure calculations ( Bandgap nature, Bandgap corresponding to GGA and HSE functional and effective mass of charge carriers) are presented. }
\centering
    \begin{tabular}{||cccccccc||}
    \hline
& \multicolumn{3}{c }{Bandgap (eV)} & \multicolumn{4}{c||}{Effective mass ($m^{*}$)} \\[1ex]
    \hline
    Structure & Bandgap nature & GGA& HSE& CBM-K & CBM-M & VBM-\(\Gamma\) & VBM-M \\[1ex]
    \hline\hline
PbI$_2$ & Indirect  & 2.58  & 3.27 & 0.26& 0.26  & 0.94 & 0.94 \\
PbBr$_2$ & Indirect& 2.82 & 3.72 & 0.33 & 0.33 & 2.96 & 2.59 \\
PbCl$_2$ & Indirect & 3.1 & 4.10  & 0.43& 0.43 & 2.03   & 1.96  \\   
PbF$_2$&Indirect &3.47&4.73&0.87&0.87&1.96&2.14\\
\hline
\end{tabular}
\end{table*}                       
\begin{table}
        \scriptsize 
\renewcommand{\arraystretch}{1}
    \begin{tabular}{ ||p{2cm}p{3cm}p{3cm}p{1.8cm}p{1.8cm}p{1.8cm}p{1.8cm}||}
    \hline
        Structure &Fermi energy(eV)& Work function(eV) & CBM (GGA)  & CBM (HSE)  & VBM (GGA) & VBM (HSE)  \\[3ex]
        \hline\hline
PbI$_2$	&-3.7&	6.1	&-5.61	&-4.47&	-7.39&	-7.73\\[1ex]
PbBr$_2$	&-4.72&	6.80	&-5.39	&-4.94	&-8.21	&-8.66\\[1ex]
PbCl$_2$	&-4.91&	7.08&	-5.53	&-5.03	&-8.63	&-9.13\\[1ex]
PbF$_2$&	-4.6&	6.69&	-4.96&	-4.33&	-8.42	&-9.05 \\[1ex]
PbIBr	&-4.18&	6.58&	-5.21&	-4.81	&-7.95	&-8.35\\[1ex]
PbICl&	-4.26&	6.77&	-5.29&	-4.94	&-8.25	&-8.6\\[1ex]
PbIF&	-4.18	&6.19	&-4.33	&-3.75	&-7.69&	-8.05\\[1ex]
PbBrCl&	-4.74	&7.14	&-5.66	&-5.2	&-8.62&	-9.08\\[1ex]
PbBrF	&-4.78&	7.22	&-5.48	&-5.05	&-8.96	&-9.39\\[1ex]
PbClF	&-4.15	&7.36&	-5.6&	-5.06	&-9.12	&-9.66\\[1ex]
\hline

    \end{tabular}
    \caption{Fermi energy, work function, VBM (GGA and HSE), CBM (GGA and HSE) which are calculated from the electrostatic potential energy calculation and band gap analysis.}
\end{table}